\documentclass[12pt]{article}
\pdfoutput=1
\usepackage{amsmath,amssymb,amsthm,amsxtra,overpic,bbm,bm,epsfig,subfigure}
\usepackage{hyperref}
\usepackage{graphicx}
\usepackage{color}
\usepackage{comment}
\usepackage{epstopdf}
\usepackage{float}
\usepackage{multirow}
\numberwithin{equation}{section}
\usepackage{cite}
\usepackage{hyperref}
\usepackage{slashed,stmaryrd}

\usepackage{lscape}%
\usepackage{array}
\usepackage{booktabs}%

\def\thefootnote{\fnsymbol{footnote}}

\begin{document}

\vspace{0.2cm}

\begin{center}
{\Large\bf Dirac neutrino mass models with a modular $S^{}_4$ symmetry}
\end{center}

\vspace{0.2cm}

\begin{center}
{\bf Xin Wang}~$^{a,~b}$~\footnote{E-mail: wangx@ihep.ac.cn}
\\
\vspace{0.2cm}
{\small
$^a$Institute of High Energy Physics, Chinese Academy of Sciences, Beijing 100049, China\\
$^b$School of Physical Sciences, University of Chinese Academy of Sciences, Beijing 100049, China}
\end{center}

\vspace{1.5cm}

\begin{abstract}
In this paper, we investigate lepton mass spectra, flavor mixing and CP violation with a modular $S^{}_4$ symmetry under the assumption that neutrinos are Dirac fermions. We find that the Majorana mass term can be forbidden by adjusting the weight of the right-handed neutrino superfields. We study the scenarios where modular forms in the neutrino sector take the lowest non-trivial weight two, while those in the charged-lepton sector vary from two to six. The corrections from renormalization-group running effects to our model have also been discussed in detail. We totally obtain twelve different classes of models and find that ten of them can accommodate current neutrino oscillation data.
\end{abstract}


\def\thefootnote{\arabic{footnote}}
\setcounter{footnote}{0}

\newpage

\section{Introduction}\label{sec:intro}
Although recent developments on solar, atmospheric, reactor and accelerator neutrino oscillation experiments have convinced us of massive neutrinos and significant lepton flavor mixing~\cite{Xing:2011zza,Tanabashi:2018oca}, it remains a mystery whether neutrinos are Dirac or Majorana particles. One can assume neutrinos to be of Majorana-type, and introduce lepton-number-violating terms in the Lagrangian. However,  since the neutrinoless double-beta decays have not yet been discovered in any related experiments, there is no obvious evidence indicating neutrinos are really Majorana fermions. Another possibility is that neutrinos are Dirac particles, which is also quite reasonable considering the other fermions we have discovered until now are all of Dirac-type. On the other hand, the Dirac neutrino assumption also embodies the famous \emph{Occam's razor}, i.e., if not necessary, we may not introduce the lepton-number-violating Majorana mass terms in the Lagrangian. 

If neutrinos are Dirac fermions, the Lagrangian is lepton number conserving. In this case, we can simply introduce three right-handed neutrinos $N^{}_{i{\rm R}}$ (for $i = 1, 2, 3$), which are singlets under the ${\rm SU}(2)_{\rm L} \times {\rm U}(1)^{}_{\rm Y}$ gauge group of the Standard Model (SM), and the gauge-invariant and lepton-number-conserving Lagrangian relevant for lepton masses and flavor mixing is written as
\begin{eqnarray}\label{eq:LagSM}
-{\cal L}^{}_{\rm Lepton} = \overline{\ell^{}_{\rm L}} Y^{}_l H E^{}_{\rm R} + \overline{\ell^{}_{\rm L}} Y^{}_\nu \widetilde{H} N^{}_{\rm R} + {\rm h.c.} \; ,
\end{eqnarray}
where $\ell^{}_{\rm L}$ and $H$ denote the left-handed lepton doublet and the Higgs doublet, $E^{}_{\rm R}$ and $N^{}_{\rm R}$ are the right-handed charged-lepton and neutrino singlets, and $Y^{}_{l}$ and $Y^{}_{\nu}$ are the charged-lepton and neutrino Yukawa coupling matrices. Note that $\widetilde{H} \equiv {\rm i} \sigma^{}_2 H^*$ has been defined in Eq.~(\ref{eq:LagSM}). After the Higgs doublet acquires its vacuum expectation value (vev), i.e., $\langle H \rangle = (0, v/\sqrt{2})^{\rm T}$ with $v \approx 246~{\rm GeV}$, the gauge symmetry is spontaneously broken down, and the charged-lepton and Dirac neutrino mass matrices are given by $M^{}_l \equiv Y^{}_l v/\sqrt{2}$ and $M^{}_{\nu} = Y^{}_\nu v/\sqrt{2}$, respectively. Then one may notice that the observed light neutrino masses ${\cal O}(M^{}_{\nu})\lesssim0.1\,{\rm eV}$ require ${\cal O}(Y^{}_{\nu}) \lesssim 10^{-12}_{}$. Such small Yukawa couplings can be generated in some models with extra spacial dimensions~\cite{Dienes:1998sb,ArkaniHamed:1998vp}, radiative mechanisms~\cite{Chang:1986bp} or gauge extensions of the SM~\cite{Ma:2014qra,Ma:2015mjd,Jana:2019mez,Jana:2019mgj}.

Non-Abelian discrete flavor symmetries have been widely applied to explain the observed lepton flavor mixing pattern~~\cite{Xing:2019vks,Altarelli:2010gt, Ishimori:2010au, King:2013eh, King:2014nza, King:2017guk,Feruglio:2019ktm}. In this framework, the Lagrangian is supposed to possess an overall discrete symmetry at some high-energy scales, and several scalar fields (i.e., flavons) are introduced into these models to break down this whole symmetry into some distinct residual symmetries in the charged-lepton and neutrino sectors. Then the flavor mixing pattern will be determined by the vev's of flavons. Recently, a new and appealing approach has been proposed in Ref.~\cite{Feruglio:2017spp}, which suggests that the modular symmetry can be utilized to account for lepton flavor mixing. Different from the traditional discrete symmetry, the Yukawa couplings in modular invariant models are regarded as modular forms with even weights, which are the functions of only one complex parameter, i.e., the modulus $\tau$. The Lagrangian keeps invariant under the finite modular symmetry group $\Gamma^{}_N$, which for a given value of $N$ is isomorphic to the non-Abelian discrete symmetry group, i.e., $\Gamma^{}_{2} \simeq S^{}_{3}$~\cite{Kobayashi:2018vbk, Kobayashi:2018wkl, Kobayashi:2019rzp, Okada:2019xqk}, $\Gamma^{}_{3} \simeq A^{}_{4}$~\cite{Kobayashi:2018scp, Criado:2018thu, deAnda:2018ecu, Okada:2018yrn, Nomura:2019jxj, Nomura:2019lnr, Ding:2019zxk, Nomura:2019yft, Okada:2019mjf, Asaka:2019vev,Okada:2020dmb}, $\Gamma^{}_{4} \simeq S^{}_4$~\cite{Penedo:2018nmg, Novichkov:2018ovf, Okada:2019lzv,Wang:2019ovr} and $\Gamma^{}_{5} \simeq A^{}_5$~\cite{Novichkov:2018nkm, Ding:2019xna, Criado:2019tzk}, and the overall symmetry will be broken once the modulus $\tau$ instead of flavons obtains its vev. Therefore, the flavon field is not necessary for the whole theory. Apart from the finite modular groups mentioned above, some other aspects of modular symmetries have also been investigated, such as the combination of modular symmetries and the CP symmetry~\cite{Novichkov:2019sqv,Kobayashi:2019uyt,Kobayashi:2020uaj}, multiple modular symmetries~\cite{deMedeirosVarzielas:2019cyj, King:2019vhv}, the double covering of modular groups~\cite{Liu:2019khw,Novichkov:2020eep,Liu:2020akv}, the $A^{}_{4}$ symmetry from the modular $S^{}_{4}$ symmetry~\cite{Kobayashi:2019mna, Kobayashi:2019xvz}, the modular residual symmetry~\cite{Novichkov:2018yse,Gui-JunDing:2019wap}, the unification of quark and lepton flavors with modular invariance~\cite{Okada:2019uoy,Okada:2020rjb}, the realization of texture zeros via the modular symmetry~\cite{Zhang:2019ngf,Lu:2019vgm}, the applications of modular symmetries on other types of seesaw models~\cite{Kobayashi:2019gtp,Nomura:2019xsb,Wang:2019xbo,Behera:2020sfe,1806045}, the origin of modular symmetries from a top-down point of view~\cite{Kobayashi:2018bff,Nilles:2020nnc,Nilles:2020kgo,Kobayashi:2020hoc,Ohki:2020bpo,Kikuchi:2020frp} and finite modular groups with higher levels~\cite{Ding:2020msi}. In addition, corrections from the K\"ahler potential to the model with a modular symmetry are discussed in Ref.~\cite{Chen:2019ewa}.

In this paper, for the first time we investigate Dirac neutrino masses, lepton flavor mixing and CP violation with a modular $S^{}_{4}$ symmetry. Thanks to the restrictions from modular weights, we find that the Majorana mass term can be forbidden in our model. We discuss the scenarios where modular forms in the neutrino sector take the lowest non-trivial weight two, while those in the charged-lepton sector vary from two to six, and obtain twelve different classes of models. Since modular symmetries intrinsically work at very high-energy scales, we take the renormalization-group (RG) running effects into consideration in order to obtain the predictions for low-energy observables. After implementing the numerical analysis, we find ten models are consistent with current neutrino oscillation data, in either the normal mass ordering (NO) case where $m^{}_1<m^{}_2<m^{}_3$ or the inverted mass ordering (IO) case where $m^{}_3<m^{}_1<m^{}_2$. In addition, we also implement the generalized CP symmetry in our model, which can make all the coupling constants in our model real and reduce the number of free parameters.

The remaining part of this paper is organized as follows. In Sec.~\ref{sec:modular}, we present a brief summary of the modular $S^{}_4$ symmetry. The Dirac neutrino mass models with a modular $S^{}_4$ symmetry are then constructed in Sec.~\ref{sec:models}. The low-energy phenomenology of lepton mass spectra, flavor mixing pattern and CP violation of these models with the corrections from RG running effects is discussed in Sec.~\ref{sec:lowphe}. Finally, we summarize our main conclusions in Sec.~\ref{sec:summary}. Some properties of the modular $S^{}_4$ symmetry group are presented in Appendix~\ref{sec:appA}.

\section{Modular $S^{}_4$ symmetry} \label{sec:modular}
In this section, we shall briefly summarize some knowledge about the modular symmetry which is helpful for this paper. As is well known, the modular symmetry is supposed to work in the supersymmetric theory, where the action ${\cal S}$ keeps invariant under the modular transformations~\cite{Feruglio:2017spp}
\begin{eqnarray}
\gamma: \tau \rightarrow \dfrac{a \tau + b}{c \tau + d} \; , \quad
\chi^{(I)}_{} \rightarrow (c \tau +d )^{-k^{}_I} \rho^{(I)}_{} (\gamma) \chi^{(I)} _{} \; ,
\label{eq:actiontransf}
\end{eqnarray}
where $\gamma$ is the element of the inhomogeneous modular group $\overline{\Gamma}$ with $a$, $b$, $c$ and $d$ being integers satisfying $ad - bc =1$, $\tau$ is an arbitrary complex number in the upper complex plane, $\rho^{(I)}(\gamma)$ denotes the representation matrix of the modular transformation $\gamma$, and $k^{}_{I}$ is the weight associated with the supermultiplet $\chi^{(I)}$. As a consequence, the K\"ahler potential ${\cal K}(\tau, \chi)$ should keep invariant up to the K\"ahler transformation, while the superpotential ${\cal W}(\tau, \chi)$ is also invariant under the modular transformation and can be expanded in terms of the supermultiplets $\chi^{(I^{}_{i})}_{}$ (for $i = 1, \cdots, n$) as
\begin{eqnarray}
{\cal W}(\tau, \chi)= \sum_{n}^{}\sum_{\{I^{}_{1},\dots,I^{}_{n}\}}^{} Y^{}_{I^{}_1\dots I^{}_n}(\tau)\chi_{}^{(I^{}_1)}\cdots\chi_{}^{(I^{}_n)} \; ,
\label{eq:surpoten}
\end{eqnarray}
where the coefficients $Y^{}_{I^{}_1\dots I^{}_n}(\tau)$ take the modular forms, which are the key elements of the modular symmetry approach. Under the finite modular group $\Gamma^{}_{N} \equiv \overline{\Gamma}/\overline{\Gamma}(N)$ (with $\overline{\Gamma}(N)$ being the principal congruence subgroup of $\overline{\Gamma}$), the modular forms transform as
\begin{eqnarray}
Y^{}_{I^{}_1\dots I^{}_n}(\tau) \rightarrow (c\tau+d)^{k^{}_Y}_{} \rho^{}_{Y} (\gamma) Y^{}_{I^{}_1 \dots I^{}_n}(\tau) \; ,
\label{eq:Yuktransf}
\end{eqnarray}
where the even integer $k^{}_{Y}$ is the weight of $Y^{}_{I^{}_1\dots I^{}_n}(\tau)$ and $\rho^{}_Y$ is the representation matrix of $\Gamma^{}_N$. $k^{}_Y$ and $\rho^{}_{Y}$ must satisfy $k^{}_{Y} = k^{}_{I^{}_1} +  \cdots + k^{}_{I^{}_N}$ and $\rho^{}_{Y} \otimes \rho^{(I^{}_{1})}_{}  \otimes \cdots \otimes \rho^{(I^{}_{N})}_{} \ni {\bf 1}$ respectively, which can be used to constrain the charge assignments of superfields and modular forms.

For the modular group $\Gamma^{}_{4} \simeq S^{}_{4}$, there are five linearly independent modular forms of the lowest non-trivial weight $k^{}_{Y}=2$, denoted as $Y^{}_i(\tau)$ for $i = 1, 2, \cdots, 5$, which form a doublet ${\bf 2}$ and a triplet ${\bf 3}^{\prime}_{}$ under the modular $S^{}_4$ symmetry transformations~\cite{Penedo:2018nmg}, namely,
\begin{eqnarray}
Y^{}_{\bf 2}(\tau) \equiv \left(\begin{matrix} Y^{}_{1}(\tau) \\ Y^{}_2 (\tau)  \end{matrix}\right) \; , \quad Y^{}_{\bf 3^{\prime}_{}} (\tau) \equiv  \left(\begin{matrix} Y^{}_{3}(\tau) \\ Y^{}_4 (\tau) \\ Y^{}_{5} (\tau) \end{matrix}\right) \; .
\label{eq:S4Y}
\end{eqnarray}
The exact expressions of $Y^{}_{i}(\tau)$ (for $i=1,2,\cdots,5$) can be found in Appendix~\ref{sec:appA}. Based on the modular forms $Y^{}_{i}(\tau)$ of weight two, one can construct the modular forms of higher weights, such as $k^{}_Y = 4$ and $k^{}_Y = 6$. For $k^{}_Y = 4$, there are totally nine independent modular forms, which transform as {\bf 1}, {\bf 2}, {\bf 3} and ${\bf 3}^{\prime}_{}$ under the modular $S^{}_{4}$ symmetry~\cite{Novichkov:2019sqv}, namely,
\begin{equation}
\begin{split}
Y^{(4)}_{\bf 1} = Y^{2}_{1}+Y^{2}_{2} \; &, \quad Y^{(4)}_{\bf 2} = \left( \begin{matrix} Y^2_2-Y^2_1 \\ 2Y^{}_1 Y^{}_2 \end{matrix} \right) \; , \\
Y^{(4)}_{\bf 3} =\left( \begin{matrix} -2Y^{}_{2}Y^{}_{3} \\ \sqrt{3}Y^{}_{1}Y^{}_{5}+Y^{}_{2}Y^{}_{4} \\ \sqrt{3}Y^{}_{1}Y^{}_{4}+Y^{}_{2}Y^{}_{5} \end{matrix} \right) \; &, \quad Y^{(4)}_{\bf 3^{\prime}_{}} =\left( \begin{matrix} 2Y^{}_{1}Y^{}_{3} \\ \sqrt{3}Y^{}_{2}Y^{}_{5} - Y^{}_{1}Y^{}_{4} \\ \sqrt{3}Y^{}_{2}Y^{}_{4} - Y^{}_{1}Y^{}_{5} \end{matrix} \right) \; ,
\end{split}
\label{eq:Y4}
\end{equation}
where the argument $\tau$ of all the modular forms is suppressed. For $k^{}_Y = 6$, we have thirteen independent modular forms, whose assignments under the $S^{}_4$ symmetry can be expressed as~\cite{Novichkov:2019sqv}
\begin{equation}
\begin{split}
Y^{(6)}_{\bf 1} = Y^{}_{1}(3Y^{2}_{2}-Y^{2}_{1}) \; &, \quad Y^{(6)}_{\bf 1^{\prime}_{}} = Y^{}_{2}(3Y^{2}_{1}-Y^{2}_{2}) \; ,\\
Y^{(6)}_{\bf 2} = (Y^{2}_{1}+Y^{2}_{2})\left( \begin{matrix} Y^{}_{1} \\ Y^{}_{2} \end{matrix} \right) \;  &, \quad Y^{(6)}_{\bf 3} =\left( \begin{matrix} Y^{}_{1}(Y^{2}_{4}-Y^{2}_{5}) \\ Y^{}_{3}(Y^{}_{1}Y^{}_{5}+\sqrt{3}Y^{}_{2}Y^{}_{4}) \\ -Y^{}_{3}(Y^{}_{1}Y^{}_{4}+\sqrt{3}Y^{}_{2}Y^{}_{5})\end{matrix} \right) \; ,\\
Y^{(6)}_{{\bf 3^{\prime}_{}},1} =(Y^{2}_{1}+Y^{2}_{2})\left( \begin{matrix} Y^{}_{3} \\ Y^{}_{4} \\ Y^{}_{5} \end{matrix} \right) \; &, \quad Y^{(6)}_{{\bf 3^{\prime}_{}},2} =\left( \begin{matrix} Y^{}_{2}(Y^{2}_{5}-Y^{2}_{4}) \\ -Y^{}_{3}(Y^{}_{2}Y^{}_{5}-\sqrt{3}Y^{}_{1}Y^{}_{4}) \\ Y^{}_{3}(Y^{}_{2}Y^{}_{4}-\sqrt{3}Y^{}_{1}Y^{}_{5}) \end{matrix} \right) \; .
\end{split}
\label{eq:Y6}
\end{equation}
\section{Dirac neutrino mass models} \label{sec:models}
Now we are going to construct the Dirac neutrino mass models with a modular $S^{}_{4}$ symmetry. Before going to the detail, we would like to make some general remarks on the model building.
\begin{itemize}
	\item One principle for the model building is that the number of free model parameters we introduce should be as small as possible. We have totally eight low-energy observables whose values are tightly restricted by current experiments: three charged-lepton masses $\{m^{}_e,m^{}_{\mu},m^{}_{\tau}\}$, two neutrino mass-squared differences $\{\Delta m^{2}_{21},\Delta m^{2}_{31}\}$ ($\{\Delta m^{2}_{21},\Delta m^{2}_{32}\}$) in the NO (IO) case and three lepton flavor mixing angles $\{\theta^{}_{12},\theta^{}_{13},\theta^{}_{23}\}$. Consequently, we should keep the number of free model parameters no more than eight in order to make our model more predictive.
	
	\item The representation of the lepton doublet superfields $\widehat{L}$ is set to the triplet {\bf 3} under the modular $S^{}_{4}$ symmetry. If doing so, the superfields for three charged-lepton singlets $\{\widehat{E}^{\rm C}_1, \widehat{E}^{\rm C}_2, \widehat{E}^{\rm C}_3\}$ should not be arranged as a triplet, since in this case, the strong mass hierarchy in the charged-lepton sector could not be easily generated. So we instead assign $\{\widehat{E}^{\rm C}_1, \widehat{E}^{\rm C}_2, \widehat{E}^{\rm C}_3\}$ to be three singlets under the group $S^{}_4$. On the other hand, the superfields for three right-handed neutrinos $\widehat{N}^{\rm C}_{}$ are also arranged to be a triplet, either {\bf 3} or ${\bf 3}^{\prime}_{}$. Note that it is not necessary to further consider $\widehat{L} \sim {\bf 3}^{\prime}_{}$, since the cases where $\{\widehat{L} \sim {\bf 3}^{\prime}_{},\widehat{N}^{\rm C} \sim {\bf 3}\}$ and $\{\widehat{L} \sim {\bf 3}^{\prime}_{},\widehat{N}^{\rm C} \sim {\bf 3}^{\prime}_{}\}$ are equivalent to those where $\{\widehat{L} \sim {\bf 3},\widehat{N}^{\rm C} \sim {\bf 3}^{\prime}_{}\}$ and $\{\widehat{L} \sim {\bf 3},\widehat{N}^{\rm C} \sim {\bf 3}\}$, respectively~\cite{Novichkov:2018ovf}. Moreover, the Higgs superfields $\{\widehat{H}^{}_{\rm u},\widehat{H}^{}_{\rm d}\}$ are set to be the trivial one dimensional irreducible representation {\bf 1} for simplicity.
	
	\item Now that the representations of $\widehat{L}$ and $\widehat{E}^{\rm C}_{i}$ (for $i=1,2,3$) have been determined, it is not difficult to find that the charged-lepton Yukawa couplings $f^{}_e(\tau)$, $f^{}_{\mu}(\tau)$ and $f^{}_{\tau}(\tau)$, which take the modular forms, should be assigned as triplets under the $S^{}_4$ symmetry to construct the modular invariant superpotentials. According to Eq.~(\ref{eq:S4Y}), there is only one ${\bf 3}^{\prime}_{}$ of weight two, so the modular forms of higher weights should be included into the model, otherwise we will be unable to obtain non-degenerate charged-lepton masses. In this paper, we assume the weight of $f^{}_{e}(\tau)$ to be two, i.e., $f^{}_{e}(\tau) \sim Y^{(2)}_{{\bf 3}^{\prime}_{}}$, while the weights of  $f^{}_{\mu}(\tau)$ and $f^{}_{\tau}(\tau)$ can vary from four to six. The Yukawa coupling coefficient $f^{}_{\rm D}(\tau)$ in the neutrino sector simply takes the modular forms of weight 2, because higher weights of $f^{}_{\rm D}(\tau)$ will inevitably bring more free parameters into the model.
\end{itemize}

Keeping the above remarks in mind, now we can write down the modular $S^{}_4$ invariant superpotential ${\cal W} = {\cal W}^{}_l+ {\cal W}^{}_{\rm D} $ in the minimal supersymmetric standard model (MSSM), where ${\cal W}^{}_{l}$ and ${\cal W}^{}_{\nu}$ are expressed as
\begin{eqnarray}
{\cal W}^{}_l = \alpha^{}_{1} \left[\left(\widehat{L} \widehat{E}^{\rm C}_{1}\right)^{}_{{\bf 3}^\prime} \left(f^{}_e(\tau)\right)^{}_{{\bf 3}^\prime}\right]^{}_{\bf 1} \widehat{H}^{}_{\rm d} &+& \alpha^{}_{2} \left[\left(\widehat{L} \widehat{E}^{\rm C}_{2}\right)^{}_{{\bf 3}^{(\prime)_{}}} \left(f^{}_{\mu}(\tau)\right)^{}_{{\bf 3}^{(\prime)_{}}} \right]^{}_{\bf 1} \widehat{H}^{}_{\rm d}  \; , \nonumber \\ 
&+& \alpha^{}_{3} \left[ \left(\widehat{L} \widehat{E}^{\rm C}_{3}\right)^{}_{{\bf 3}^{(\prime)_{}}} \left(f^{}_{\tau}(\tau)\right)^{}_{{\bf 3}^{(\prime)_{}}} \right]^{}_{\bf 1}  \widehat{H}^{}_{\rm d} \nonumber \; ,\\
{\cal W}^{}_{\rm D} = g^{}_{1}\left[\left(\widehat{L} \widehat{N}^{\rm C}_{}\right)^{}_{\bf 3} \left(f^{}_{\rm D}(\tau) \right)^{}_{\bf 3}\right]^{}_{\bf 1} \widehat{H}^{}_{\rm u} &+& g^{}_{2}\left[\left( \widehat{L} \widehat{N}^{\rm C}_{}\right)^{}_{{\bf 3}^{\prime}_{}} \left(f^{\prime}_{\rm D}(\tau) \right)^{}_{{\bf 3}^{\prime}_{}}\right]^{}_{\bf 1} \widehat{H}^{}_{\rm u}\; ,
\label{eq:superp1}
\end{eqnarray}
where $\alpha^{}_{1}$, $\alpha^{}_{2}$ and $\alpha^{}_{3}$ are three real and positive parameters in the charged-lepton sector, $g^{}_{1}$ and $g^{}_{2}$ are two coefficients in the neutrino sector and $f^{\prime}_{\rm D}(\tau)$ is another neutrino Yukawa coupling coefficient distinct from $f^{}_{\rm D}(\tau)$. As has been mentioned above, the invariance of ${\cal W}$ under the modular transformation requires $k^{}_{Y} = k^{}_{I^{}_1} +  \cdots + k^{}_{I^{}_N}$ holds for each term in Eq.~(\ref{eq:superp1}). To be specific, we have the following relations
\begin{eqnarray}
k^{}_{L}+k^{}_{E^{}_1} = k^{}_{e} \; , \quad k^{}_{L}+k^{}_{E^{}_2} = k^{}_{\mu} \; ,\quad k^{}_{L}+k^{}_{E^{}_3} = k^{}_{\tau} \; ,\quad k^{}_{L}+k^{}_{N} = k^{(\prime)}_{\rm D} \; ,
\label{eq:weighteq}
\end{eqnarray}
where $-k^{}_{L}$, $-k^{}_{E^{}_1}$, $-k^{}_{E^{}_2}$, $-k^{}_{E^{}_3}$ and $-k^{}_{N}$ are the weights of the corresponding superfields $\widehat{L}$, $\widehat{E}^{\rm C}_{1}$, $\widehat{E}^{\rm C}_{2}$, $\widehat{E}^{\rm C}_{3}$ and $\widehat{N}^{\rm C}_{}$ while $k^{}_{e}$, $k^{}_{\mu}$, $k^{}_{\tau}$, $k^{}_{\rm D}$ and $k^{\prime}_{\rm D}$ are the weights of $f^{}_e(\tau)$, $f^{}_{\mu}(\tau)$, $f^{}_{\tau}(\tau)$, $f^{}_{\rm D}(\tau)$ and $f^{\prime}_{\rm D}(\tau)$. Note that we have five superfields but only four equations in Eq.~(\ref{eq:weighteq}), hence there is some freedom to adjust the weights of these superfields. The Dirac nature of neutrinos in our model indicates that the Majorana mass term proportional to $\widehat{N}^{\rm C}_{}\widehat{N}^{\rm C}_{}$ should not exist. Since the modular weights are non-negative integers, if $k^{}_{N}$ is required to be a positive integer, we could not find proper modular forms with  weight $k$ that satisfies $2k^{}_{N}+k=0$. Therefore the Majorana mass term will be automatically forbidden and we do not need to introduce extra ${\rm U(1)}$ global symmetries.

When the modular symmetry is broken, the charged-lepton Yukawa coupling matrix $\lambda^{}_{l}$ and the Dirac neutrino Yukawa coupling matrix $\lambda^{}_{\rm D}$ can be generated and the superpotential ${\cal W}$ turns out to be
\begin{eqnarray}
{\cal W} = \lambda^{}_{l} \widehat{L} \widehat{H}^{}_{{\rm d}} \widehat{E}^{\rm C}_{}   + \lambda^{}_{\rm D}  \widehat{L} \widehat{H}^{}_{\rm u} \widehat{N}^{\rm C}  \; .
\label{eq:superp2}
\end{eqnarray}
After the supersymmetry breaking and the ${\rm SU(2)}^{}_{\rm L}\times{\rm U(1)}^{}_{\rm Y}$ gauge symmetry breakdown, all the Higgs fields get their own vev's, and one can then obtain the lepton mass matrices $M^{}_l$ and $M^{}_{\nu}$ as
\begin{eqnarray}
M^{}_l = v^{}_{\rm d} \lambda^{\ast}_l/\sqrt{2} \; , \quad  M^{}_{\nu} = v^{}_{\rm u} \lambda^{\ast}_{\rm D}/\sqrt{2} \; , \label{eq:Mlambda}
\end{eqnarray} 
where $v^{}_{\rm d} = v \cos\beta$ and $v^{}_{\rm u} = v\sin\beta$ are respectively the vev's of the neutral scalar component fields of $\widehat{H}^{}_{\rm d}$ and $\widehat{H}^{}_{\rm u}$, with $\tan\beta \equiv v^{}_{\rm u}/v^{}_{\rm d}$ being their ratio. Note that here we use ``$\ast$'' to denote the complex conjugation and the left-right convention for the fermion mass terms is adopted in this paper. Then, by using the product rules of the modular $S^{}_{4}$ symmetry presented in Appendix~\ref{sec:appA}, we can construct the mass matrices for charged leptons as well as neutrinos.
\subsection{Charged-lepton mass matrices}
In the charged-lepton sector we set $f^{}_e(\tau)$ to be $Y^{(2)}_{{\bf 3}^{\prime}_{}}$, while $f^{}_\mu(\tau)$ and $f^{}_\tau(\tau)$ can take the forms among $Y^{(4)}_{\bf 3}$, $Y^{(4)}_{{\bf 3}^{\prime}_{}}$, $Y^{(6)}_{\bf 3}$ and $Y^{(6)}_{{\bf 3}^{\prime}_{},2}$. Actually there is another distinct form of ${\bf 3}^{\prime}_{}$ of weight 6, $Y^{(6)}_{{\bf 3}^{\prime}_{},1}$, which is proportional to $Y^{(2)}_{{\bf 3}^{\prime}_{}}$. If both $Y^{(2)}_{{\bf 3}^{\prime}_{}}$ and $Y^{(6)}_{{\bf 3}^{\prime}_{},1}$ enter the charged-lepton mass matrix, two of the mass eigenvalues will be degenerate. Therefore, we do not take $Y^{(6)}_{{\bf 3}^{\prime}_{},1}$ into consideration. In the following, we list six different kinds of charged-lepton mass matrices obtained by adjusting the forms of $f^{}_\mu(\tau)$ and $f^{}_\tau(\tau)$, which are labeled by {\bf L1} --- {\bf L6}.
\begin{itemize}
	\item {\bf L1}: $f^{}_{\mu}(\tau)\sim Y^{(4)}_{\bf 3}$, $f^{}_{\tau}(\tau)\sim Y^{(4)}_{{\bf 3}^{\prime}_{}}$
	\begin{eqnarray}
	M^{}_l = \frac{v^{}_{\rm d}}{\sqrt{2}} \left( \begin{matrix}
	\alpha^{}_{1} Y^{}_{3} &&  -2\alpha^{}_{2} Y^{}_{2}Y^{}_{3}   &&  2\alpha^{}_{3} Y^{}_{1}Y^{}_{3}   \\ \alpha^{}_{1} Y^{}_{5}
	&& \alpha^{}_{2} (\sqrt{3}Y^{}_{1}Y^{}_{4}+Y^{}_{2}Y^{}_{5})   && \alpha^{}_{3} (\sqrt{3}Y^{}_{2}Y^{}_{4}-Y^{}_{1}Y^{}_{5}) \\ \alpha^{}_{1} Y^{}_{4}
	&&  \alpha^{}_{2} (\sqrt{3}Y^{}_{1}Y^{}_{5}+Y^{}_{2}Y^{}_{4})   && \alpha^{}_{3} (\sqrt{3}Y^{}_{2}Y^{}_{5}-Y^{}_{1}Y^{}_{4})
	\end{matrix} \right)^* \; ;
	\label{eq:Me1}
	\end{eqnarray}
	
	\item {\bf L2}: $f^{}_{\mu}(\tau)\sim Y^{(4)}_{\bf 3}$, $f^{}_{\tau}(\tau)\sim Y^{(6)}_{\bf 3}$
	\begin{eqnarray}
	M^{}_l = \frac{v^{}_{\rm d}}{\sqrt{2}} \left( \begin{matrix}
	\alpha^{}_{1} Y^{}_{3} &&  -2\alpha^{}_{2} Y^{}_{2}Y^{}_{3}   &&  \alpha^{}_{3} Y^{}_{1}(Y^{2}_{4}-Y^{2}_{5})   \\ \alpha^{}_{1} Y^{}_{5}
	&& \alpha^{}_{2} (\sqrt{3}Y^{}_{1}Y^{}_{4}+Y^{}_{2}Y^{}_{5})   && -\alpha^{}_{3} Y^{}_{3}(Y^{}_{1}Y^{}_{4}+\sqrt{3}Y^{}_{2}Y^{}_{5}) \\ \alpha^{}_{1} Y^{}_{4}
	&&  \alpha^{}_{2} (\sqrt{3}Y^{}_{1}Y^{}_{5}+Y^{}_{2}Y^{}_{4})   && \alpha^{}_{3} Y^{}_{3}(Y^{}_{1}Y^{}_{5}+\sqrt{3}Y^{}_{2}Y^{}_{4})
	\end{matrix} \right)^* \; ;
	\label{eq:Me2}
	\end{eqnarray}
	
	\item {\bf L3}: $f^{}_{\mu}(\tau)\sim Y^{(4)}_{\bf 3}$, $f^{}_{\tau}(\tau)\sim Y^{(6)}_{{\bf 3}^{\prime}_{},2}$
	\begin{eqnarray}
	M^{}_l = \frac{v^{}_{\rm d}}{\sqrt{2}} \left( \begin{matrix}
	\alpha^{}_{1} Y^{}_{3} &&  -2\alpha^{}_{2} Y^{}_{2}Y^{}_{3}   &&  \alpha^{}_{3} Y^{}_{2}(Y^{2}_{5}-Y^{2}_{4})   \\ \alpha^{}_{1} Y^{}_{5}
	&& \alpha^{}_{2} (\sqrt{3}Y^{}_{1}Y^{}_{4}+Y^{}_{2}Y^{}_{5})   && \alpha^{}_{3} Y^{}_{3}(Y^{}_{2}Y^{}_{4}-\sqrt{3}Y^{}_{1}Y^{}_{5}) \\ \alpha^{}_{1} Y^{}_{4}
	&&  \alpha^{}_{2} (\sqrt{3}Y^{}_{1}Y^{}_{5}+Y^{}_{2}Y^{}_{4})   && -\alpha^{}_{3} Y^{}_{3}(Y^{}_{2}Y^{}_{5}-\sqrt{3}Y^{}_{1}Y^{}_{4})
	\end{matrix} \right)^* \; ;
	\label{eq:Me3}
	\end{eqnarray}
	
	\item {\bf L4}: $f^{}_{\mu}(\tau)\sim Y^{(4)}_{{\bf 3}^{\prime}_{}}$, $f^{}_{\tau}(\tau)\sim Y^{(6)}_{\bf 3}$
	\begin{eqnarray}
	M^{}_l = \frac{v^{}_{\rm d}}{\sqrt{2}} \left( \begin{matrix}
	\alpha^{}_{1} Y^{}_{3} &&  2\alpha^{}_{2} Y^{}_{1}Y^{}_{3}   &&  \alpha^{}_{3} Y^{}_{1}(Y^{2}_{4}-Y^{2}_{5})   \\ \alpha^{}_{1} Y^{}_{5}
	&& \alpha^{}_{2} (\sqrt{3}Y^{}_{2}Y^{}_{4}-Y^{}_{1}Y^{}_{5})   && -\alpha^{}_{3} Y^{}_{3}(Y^{}_{1}Y^{}_{4}+\sqrt{3}Y^{}_{2}Y^{}_{5}) \\ \alpha^{}_{1} Y^{}_{4}
	&&  \alpha^{}_{2} (\sqrt{3}Y^{}_{2}Y^{}_{5}-Y^{}_{1}Y^{}_{4})   && \alpha^{}_{3} Y^{}_{3}(Y^{}_{1}Y^{}_{5}+\sqrt{3}Y^{}_{2}Y^{}_{4})
	\end{matrix} \right)^* \; ;
	\label{eq:Me4}
	\end{eqnarray}
	
	\item {\bf L5}: $f^{}_{\mu}(\tau)\sim Y^{(4)}_{{\bf 3}^{\prime}_{}}$, $f^{}_{\tau}(\tau)\sim Y^{(6)}_{{\bf 3}^{\prime}_{},2}$
	\begin{eqnarray}
	M^{}_l = \frac{v^{}_{\rm d}}{\sqrt{2}} \left( \begin{matrix}
	\alpha^{}_{1} Y^{}_{3} &&  2\alpha^{}_{2} Y^{}_{1}Y^{}_{3}   &&  \alpha^{}_{3} Y^{}_{2}(Y^{2}_{5}-Y^{2}_{4})   \\ \alpha^{}_{1} Y^{}_{5}
	&& \alpha^{}_{2} (\sqrt{3}Y^{}_{2}Y^{}_{4}-Y^{}_{1}Y^{}_{5})   && \alpha^{}_{3} Y^{}_{3}(Y^{}_{2}Y^{}_{4}-\sqrt{3}Y^{}_{1}Y^{}_{5}) \\ \alpha^{}_{1} Y^{}_{4}
	&&  \alpha^{}_{2} (\sqrt{3}Y^{}_{2}Y^{}_{5}-Y^{}_{1}Y^{}_{4})   && -\alpha^{}_{3} Y^{}_{3}(Y^{}_{2}Y^{}_{5}-\sqrt{3}Y^{}_{1}Y^{}_{4})
	\end{matrix} \right)^* \; ;
	\label{eq:Me5}
	\end{eqnarray}
	
	\item {\bf L6}: $f^{}_{\mu}(\tau)\sim Y^{(6)}_{\bf 3}$, $f^{}_{\tau}(\tau)\sim Y^{(6)}_{{\bf 3}^{\prime}_{},2}$
	\begin{eqnarray}
	M^{}_l = \frac{v^{}_{\rm d}}{\sqrt{2}} \left( \begin{matrix}
	\alpha^{}_{1} Y^{}_{3} &&  \alpha^{}_{2} Y^{}_{1}(Y^{2}_{4}-Y^{2}_{5})   &&  \alpha^{}_{3} Y^{}_{2}(Y^{2}_{5}-Y^{2}_{4})   \\ \alpha^{}_{1} Y^{}_{5}
	&& -\alpha^{}_{2} Y^{}_{3}(Y^{}_{1}Y^{}_{4}+\sqrt{3}Y^{}_{2}Y^{}_{5})    && \alpha^{}_{3} Y^{}_{3}(Y^{}_{2}Y^{}_{4}-\sqrt{3}Y^{}_{1}Y^{}_{5}) \\ \alpha^{}_{1} Y^{}_{4}
	&&  \alpha^{}_{2} Y^{}_{3}(Y^{}_{1}Y^{}_{5}+\sqrt{3}Y^{}_{2}Y^{}_{4})  && -\alpha^{}_{3} Y^{}_{3}(Y^{}_{2}Y^{}_{5}-\sqrt{3}Y^{}_{1}Y^{}_{4})
	\end{matrix} \right)^* \; .
	\label{eq:Me6}
	\end{eqnarray}
\end{itemize}

Actually one can always exchange the modular forms of $f^{}_{e}(\tau)$, $f^{}_{\mu}(\tau)$ and $f^{}_{\tau}(\tau)$, and get new charged-lepton mass matrices. However, they differ from the above six matrices only by the permutations of columns, which will not affect the forms of $M^{}_l M^{\dag}_l$~\cite{Novichkov:2018ovf}.
\subsection{Neutrino mass matrices}
In the neutrino sector, the superfields $\widehat{N}^{\rm C}_{}$ are set to be either {\bf 3} or ${\bf 3}^{\prime}_{}$ under the modular $S^{}_{4}$ symmetry and the Yukawa couplings are with a weight of 2. Then we can obtain two different neutrino mass matrices, labeled by {\bf N1} and {\bf N2}.
\begin{itemize}
	\item {\bf N1}: $\widehat{N}^{\rm C}_{} \sim {\bf 3}$
		\begin{eqnarray}
	M^{}_{\nu} =  \frac{v^{}_{\rm u}}{\sqrt{2}}  \left[ g^{}_{1}\left(
	\begin{matrix}
	Y^{}_{1} && 0 && 0 \\ 
	0  && \dfrac{\sqrt{3}}{2}Y^{}_{2} &&  -\dfrac{1}{2}Y^{}_{1} \\ 
	0 && -\dfrac{1}{2}Y^{}_{1} && \dfrac{\sqrt{3}}{2}Y^{}_{2}
	\end{matrix}\right) + g^{}_{2} \left(
	\begin{matrix}
	0  && Y^{}_5 && -Y^{}_{4} \\ 
	-Y^{}_5 && 0  &&  Y^{}_3   \\
	Y^{}_4 && -Y^{}_3  && 0 \end{matrix}\right) \right]^* \; ;
	\label{eq:MD1}
	\end{eqnarray}
	
	\item {\bf N2}: $\widehat{N}^{\rm C}_{} \sim {\bf 3}^{\prime}_{}$
	\begin{eqnarray}
	M^{}_{\nu} =  \frac{v^{}_{\rm u}}{\sqrt{2}}  \left[ g^{}_{1}\left(
	\begin{matrix}
	-Y^{}_{2} && 0 && 0 \\ 
	0  && \dfrac{\sqrt{3}}{2}Y^{}_{1} &&  \dfrac{1}{2}Y^{}_{2} \\ 
	0 && \dfrac{1}{2}Y^{}_{2} && \dfrac{\sqrt{3}}{2}Y^{}_{1}
	\end{matrix}\right) + g^{}_{2} \left(
	\begin{matrix}
	0  && -Y^{}_4 && Y^{}_{5} \\ 
	-Y^{}_4 && -Y^{}_3  &&  0   \\
	Y^{}_5 && 0  && Y^{}_3 \end{matrix}\right) \right]^* \; .
	\label{eq:MD2}
	\end{eqnarray}
	
\end{itemize}
Note that we can assume the coefficient $g^{}_1$ to be real and extract it out of the square brackets in Eqs.~(\ref{eq:MD1})-(\ref{eq:MD2}) without loss of generality. Then it is convenient to parametrize the other complex parameter as $g^{}_{2}/g^{}_{1} \equiv \widetilde{g} = g {\rm e}^{{\rm i}\phi^{}_{g}}_{}$ with $g = |\widetilde{g}|$ and $\phi^{}_g \equiv \arg(\widetilde{g})$. Combining the mass matrices in the charged-lepton and neutrino sectors, we can finally attain twelve different Dirac neutrino mass models labeled by {\bf L1N1} --- {\bf L6N2}. As we can see, Totally eight real parameters are introduced into these twelve models, which are  $\{{\rm Re}\,\tau, {\rm Im}\,\tau \}$ related to the modulus $\tau$, $\{\alpha^{}_1, \alpha^{}_2 , \alpha^{}_3 \}$ in the charged-lepton sector and  $\{g, \phi^{}_g, v^{}_{\rm u} g^{}_1 / \sqrt{2}\}$ in the neutrino sector.

\subsection{Models with generalized CP symmetry}
The combination of the generalized CP (gCP) symmetry and the modular symmetry~\cite{Novichkov:2019sqv} can reduce the number of free model parameters and enhance the predictive power of our model. a gCP transformation acting on the chiral supermultiplet $\chi^{(I)}_{}$ is defined as
\begin{eqnarray}
\chi^{(I)}_{}(x) \stackrel{\rm CP}{\longrightarrow} X^{}_{\bf r} \overline{\chi}^{(I)}_{}(x^{}_{\rm P}) \; ,
\label{eq:gcpdef}
\end{eqnarray}
where $\overline{\chi}^{(I)}_{}(x^{}_{\rm P})$ denotes the conjugate superfield with $x^{}_{\rm P} = (t,-\vec{x})$ and $X^{}_{\bf r}$ represents a unitary matrix acting on the flavor space. According to Ref.~\cite{Novichkov:2019sqv}, the modulus $\tau$ transforms under CP as
\begin{eqnarray}
\tau \stackrel{\rm CP}{\longrightarrow} -\tau^{\ast}_{} \; .
\label{eq:tauCP}
\end{eqnarray}
The requirement that the subsequent action of the CP, modular and inverse CP transformations should be represented by another element of the modular group implies
\begin{eqnarray}
X^{}_{\bf{r}} \rho_{\bf{r}}^{*}(\gamma) X_{\bf{r}}^{-1}=\rho_{\bf{r}}^{}(u(\gamma)) \; ,
\label{eq:consist}
\end{eqnarray}
where $u(\gamma)$ is an outer automorphism of the modular group. Eq.~(\ref{eq:consist}) should be satisfied for all the elements in the finite modular group $\Gamma^{}_N$, therefore we can determine the form of $X^{}_{\bf r}$ by solving Eq.~(\ref{eq:consist}). It turns out that $X^{}_{\bf r} = \mathbb{I}$ with $\mathbb{I}$ being the identity element if the representation matrices of both $S$ and $T$ are symmetric~\cite{Novichkov:2019sqv}, which is just the basis selected in this paper. Furthermore, as can be seen in Appendix~\ref{sec:appA}, all the Clebsch-Gordan coefficients are real, therefore the modular forms $Y^{(k)}_{\bf r}$ will transform under CP as
\begin{eqnarray}
Y_{\bf{r}}^{(k)}(\tau) \stackrel{\mathrm{C P}}{\longrightarrow} Y_{\bf{r}}^{(k)}\left(-\tau^{*}\right)=\left[Y_{\bf{r}}^{(k)}(\tau)\right]^{*} \; .
\label{eq:modformCP}
\end{eqnarray}
Then one can prove that the gCP symmetry requires all the coupling parameters in our model to be real. To be specific, the parameter $\widetilde{g}$, which is originally a complex number, should be real after considering the gCP symmetry. Therefore, the number of free parameters in our model will reduce to seven.

\subsection{Renormalization-group running effects}
The modular symmetry usually works at a very high-energy scale $\Lambda$. However, the oscillation parameters are measured at the electroweak scale which is characterized by the mass of the $Z$ gauge boson $m^{}_{Z} \sim 91.2~{\rm GeV}$. Therefore, in order to obtain the accurate predictions for low-energy observables, we should also include the radiative corrections to flavor mixing parameters in our model via the renormalization-group (RG) equations, which could be important especially for large values of $\tan \beta$ or nearly-degenerate neutrino masses~\cite{Ohlsson:2013xva}. Actually the corrections from RG running effects to models with the modular symmetry have been discussed in Refs.~\cite{Criado:2018thu,Wang:2019ovr}. In this paper, the radiative corrections are directly embedded in our analysis. We assume the modular symmetry is working at the grand unified theories (GUT) scale, where $\Lambda=\Lambda^{}_{\rm GUT} = 2 \times 10^{16}_{}~{\rm GeV}$, and the predictions for oscillation parameters at the electroweak scale are obtained after the RG running.

Without loss of generality, we can work in the basis where the charged-lepton Yukawa coupling matrix $\widetilde{Y}^{}_l \equiv {\rm Diag}\{y^{}_e, y^{}_\mu, y^{}_\tau\}$ with $y^{}_\alpha = \sqrt{2} m^{}_\alpha/v^{}_{\rm d}$ (for $\alpha = e, \mu, \tau$) is diagonal, then the Dirac neutrino Yukawa matrix will be $\widetilde{Y}^{}_\nu = U^\dagger_l Y^{}_\nu$, where $U^{}_l$ is a unitary matrix diagonalizing $Y^{}_l$ via $U^\dagger_l Y^{}_l Y^\dagger_l U^{}_l = \widetilde{Y}^2_l$. In this basis, the one-loop RG equations of $\widetilde{Y}^{}_{\nu}$ and $\widetilde{Y}^{}_{l}$ can be written as~\cite{Cheng:1973nv,Machacek:1983fi}
\begin{eqnarray}
16\pi^2_{} \frac{{\rm d}\widetilde{Y}^{}_{\nu}}{{\rm d}t} &=& \left[\alpha^{}_{\nu}+C^{\nu}_{\nu}\left(\widetilde{Y}^{}_{\nu}\widetilde{Y}^{\dag}_{\nu}\right)+C^{l}_{\nu}\left(\widetilde{Y}^{}_{l}\widetilde{Y}^{\dag}_{l}\right)\right]\widetilde{Y}^{}_{\nu} \; , \label{eq:RGE1}\\
16\pi^2_{} \frac{{\rm d}\widetilde{Y}^{}_{l}}{{\rm d}t} &=& \left[\alpha^{}_{l}+C^{\nu}_{l}\left(\widetilde{Y}^{}_{\nu}\widetilde{Y}^{\dag}_{\nu}\right)+C^{l}_{l}\left(\widetilde{Y}^{}_{l}\widetilde{Y}^{\dag}_{l}\right)\right]\widetilde{Y}^{}_{l} \; , \label{eq:RGE2}
\end{eqnarray}
where $t \equiv \ln (\mu/\Lambda^{}_{\rm GUT})$ with $\mu$ being the renormalization scale. In the framework of the MSSM
\begin{eqnarray}
C^{\nu}_{\nu} = C^{l}_{l} = + 3 \; , \quad C^{l}_{\nu} = C^{\nu}_{l} =  +1 \; ,
\label{eq:C_MSSM}
\end{eqnarray}
and 
\begin{eqnarray}
\alpha^{}_{\nu} &=& -\frac{3}{5}\mathfrak{g}^2_1 -3\mathfrak{g}^2_2 + {\rm Tr}\left[3(Y^{}_{\rm u}Y^{\dag}_{\rm u})+\widetilde{Y}^{}_{\nu}\widetilde{Y}^{\dag}_{\nu}\right] \; , \nonumber \\
\alpha^{}_{l} &=& -\frac{9}{5}\mathfrak{g}^2_1 -3\mathfrak{g}^2_2 + {\rm Tr}\left[\widetilde{Y}^{}_{\nu}\widetilde{Y}^{\dag}_{\nu}+3\left(\widetilde{Y}^{}_{l}\widetilde{Y}^{\dag}_{l}\right)\right] \; ,
\label{RGEco_MSSM}
\end{eqnarray}
where $\mathfrak{g}^{}_1$ and $\mathfrak{g}^{}_2$ denote respectively the ${\rm SU(2)}^{}_{\rm L}$ and ${\rm U(1)}^{}_{\rm Y}$ gauge couplings, and $Y^{}_{\rm u}$ is the up-type quark Yukawa coupling matrix. The supersymmetry can be broken down at the energy scale $m^{}_{\rm SUSY}$ above the electroweak scale, where we have the following tree-level matching conditions~\cite{Criado:2018thu}
\begin{eqnarray}
\left(\widetilde{Y}^{}_{l}\right)^{}_{\rm SM}(m^{}_{\rm SUSY}) &=& \left(\widetilde{Y}^{}_{l}\right)^{}_{\rm MSSM}(m^{}_{\rm SUSY})\cos \beta \; , \nonumber \label{eq:matchingYl} \\
\left(\widetilde{Y}^{}_{\nu}\right)^{}_{\rm SM}(m^{}_{\rm SUSY}) &=& \left(\widetilde{Y}^{}_{\nu}\right)^{}_{\rm MSSM}(m^{}_{\rm SUSY})\sin \beta \; . \label{eq:matchingYnu} 
\end{eqnarray}
The RG running should be implemented in the framework of the SM below the energy scale $m^{}_{\rm SUSY}$. In the SM, we have 
\begin{eqnarray}
C^{\nu}_{\nu} = C^{l}_{l} = + \frac{3}{2} \; , \quad C^{l}_{\nu} = C^{\nu}_{l} =  -\frac{3}{2} \; ,
\label{eq:C_SM}
\end{eqnarray}
and
\begin{eqnarray}
\alpha^{}_{\nu} &=& -\frac{9}{20}\mathfrak{g}^2_1 -\frac{9}{4}\mathfrak{g}^2_2 + {\rm Tr}\left[3(Y^{}_{\rm u}Y^{\dag}_{\rm u})+3(Y^{}_{\rm d}Y^{\dag}_{\rm d})+\widetilde{Y}^{}_{\nu}\widetilde{Y}^{\dag}_{\nu}+\widetilde{Y}^{}_{l}\widetilde{Y}^{\dag}_{l}\right] \; , \nonumber \\
\alpha^{}_{l} &=& -\frac{9}{4}\mathfrak{g}^2_1 -\frac{9}{4}\mathfrak{g}^2_2 + {\rm Tr}\left[3(Y^{}_{\rm u}Y^{\dag}_{\rm u})+3(Y^{}_{\rm d}Y^{\dag}_{\rm d})+\widetilde{Y}^{}_{\nu}\widetilde{Y}^{\dag}_{\nu}+\widetilde{Y}^{}_{l}\widetilde{Y}^{\dag}_{l}\right] \; ,
\label{RGEco_SM}
\end{eqnarray}
where $Y^{}_{\rm d}$ is the down-type quark Yukawa coupling matrix.

Eq.~(\ref{eq:RGE2}) indicates that the charged-lepton Yukawa coupling matrix will keep diagonal during the RG running. Therefore the RG equation of $\widetilde{Y}^{}_l$ will be reduced to three individual equations corresponding to three diagonal elements of $\widetilde{Y}^{}_l$, which can be easily solved out. Then the approximate solution to Eq.~(\ref{eq:RGE1}) turns out to be
\begin{eqnarray}
\widetilde{Y}^{}_{\nu}(m^{}_Z) = I^{}_{\nu} 
\left(\begin{matrix}
I^{}_e && 0 && 0 \\
0 && I^{}_{\mu} && 0 \\
0 && 0 && I^{}_{\tau}
\end{matrix}\right)
 \widetilde{Y}^{}_{\nu}(\Lambda^{}_{\rm GUT}) \; ,
\label{eq:Ynusol}
\end{eqnarray}
where $I^{}_{\nu}$ and $I^{}_{\alpha}$ (for $\alpha=e,\mu,\tau$) are the evolution functions defined as
\begin{eqnarray}
I^{}_\nu & = & {\rm exp}\left[-\frac{1}{16\pi^2_{}}\int^{\ln (\Lambda^{}_{\rm GUT}/m^{}_Z)}_{0}\alpha^{}_{\nu}(t) {\rm d}t\right] \; , \nonumber \\
I^{}_\alpha & = & {\rm exp}\left[-\frac{1}{16\pi^2_{}}\int^{\ln (\Lambda^{}_{\rm GUT}/m^{}_Z)}_{0} y^2_\alpha (t) {\rm d}t\right] \; . \label{eq:evofun}
\end{eqnarray}
One can observe from Eq.~(\ref{eq:evofun}) that $I^{}_{\nu}$ only affects the absolute mass scale of neutrinos while  $I^{}_{\alpha}$ will modify all the oscillation parameters. Since the values of $y^{}_e$ and $y^{}_{\mu}$ are extremely small, $I^{}_e \approx I^{}_{\mu} \approx 1$ can be regarded as a good approximation. Then $I^{}_{\tau}$ makes the dominant contributions to the radiative corrections of $\widetilde{Y}^{}_{\nu}$. After the model parameters are fixed, we can immediately obtain $\widetilde{Y}^{}_{\nu}(\Lambda^{}_{\rm GUT})$, which can be regarded as the initial conditions at $\Lambda^{}_{\rm GUT}$. Then by using Eq.~(\ref{eq:Ynusol}), we calculate $\widetilde{Y}^{}_{\nu}(m^{}_Z)$ at the electroweak scale via the RG running, from which one can determine the values of oscillation parameters. The detailed numerical analysis will be presented in next section.

\section{Numerical analysis} \label{sec:lowphe}
\subsection{The strategy}
After establishing the concrete models, we implement the numerical analysis in this section. In order to find out the feasible parameter space, we calculate the predictions for neutrino oscillation parameters in our model, and then compare them with the global-fit results from NuFIT 4.1~\cite{Esteban:2018azc} without including the atmospheric neutrino data from Super-Kamiokande, which are listed in Table~\ref{table:gfit}. Note that the current constraint on the Dirac CP-violating phase $\delta$ from the global-fit results is rather weak, therefore we will not include the information of $\delta$. The detailed numerical analysis strategy is listed as follows.

\begin{itemize}
	\item First of all, as has been clearly explained in Ref.\cite{Novichkov:2018ovf}, it is sufficient to scan the parameter $\tau$ in the fundamental domain defined as
	\begin{eqnarray}
	{\cal G} = \left\{ \tau \in \mathbb{C}: \quad {\rm Im}\,\tau > 0, \; | {\rm Re}\,\tau| \leq 0.5, \; |\tau| \geq 1 \right\} \; .
	\label{eq:fundo}
	\end{eqnarray}
	While in this paper, the scan range of ${\rm Re}\,\tau$ is further restricted to $0 \leq {\rm Re}\,\tau \leq 0.5$. The numerical results for the conjugate range where  $-0.5 \leq {\rm Re}\,\tau \leq 0$ can be easily obtained by reversing the sign of the Dirac CP-violating phase $\delta$. If the values of ${\rm Re}\,\tau$ and ${\rm Im}\,\tau$ are given, the parameters $\alpha^{}_1$, $\alpha^{}_2$ and $\alpha^{}_3$ in the charged-lepton sector can be numerically calculated via the following equations
	\begin{eqnarray}
	{\rm Tr} \left( M^{}_l M^{\dag}_l \right) &=& m^{2}_{e} + m^{2}_{\mu} + m^{2}_{\tau} \; ,  \label{eq:tr}\\
	{\rm Det}\left( M^{}_l M^{\dag}_l \right) &=& m^{2}_{e} m^{2}_{\mu} m^{2}_{\tau} \; , \label{eq:det}\\
	\dfrac{1}{2}\left[{\rm Tr} \left(M^{}_l M^{\dag}_l\right)\right]^2_{} - \dfrac{1}{2}{\rm Tr}\left[ (M^{}_l M^{\dag}_l)^2_{}\right] &= & m^{2}_{e}m^{2}_{\mu}+m^{2}_{\mu}m^{2}_{\tau}+m^{2}_{\tau}m^{2}_{e} \; , \label{eq:tr2}
	\end{eqnarray}
    where we use the best-fit values of charged-lepton masses at the GUT scale with $m^{}_{\rm SUSY}=10~{\rm TeV}$ and $\tan\beta=10$~\cite{Antusch:2013jca,Lu:2019vgm}, which are $m^{}_e = 0.510~{\rm MeV}$, $m^{}_{\mu} = 107.8~{\rm MeV}$ and $m^{}_{\tau} = 1840.14~{\rm MeV}$. At the same time, the charged-lepton mass matrix $M^{}_l$ at the GUT scale can also be determined, so as the unitary matrix $U^{}_{l}$. 
	
	\item Next we randomly generate the values of $g$ and $\phi^{}_g$ in the region where $g \in [10^{-4}_{},10^4]$ and $\phi^{}_g \in [-\pi,\pi]$. After $g$ and $\phi^{}_g$ are fixed, the neutrino mass matrix $\widetilde{M}^{}_{\nu}=v^{}_{\rm u}U^{\dag}_{l}Y^{}_{\nu}/\sqrt{2}$ in the basis where the charged-lepton mass matrix is diagonal can be determined up to an overall factor $v^{}_{\rm u}g^{}_1/\sqrt{2}$ at the GUT scale. Assuming $m^{}_{\rm SUSY}= 10~{\rm TeV}$ and $\tan\beta=10$, we can then obtain the corresponding $\widetilde{M}^{}_{\nu}$ at the electroweak scale by solving the RG equations. We define a ratio $r \equiv \Delta m^{2}_{21}/\Delta m^{2}_{31}$ ($\Delta m^{2}_{21}/|\Delta m^{2}_{32}|$) in the NO (IO) case. Using the global-fit results of $\Delta m^{2}_{21}$ and $\Delta m^{2}_{31}$ ($\Delta m^{2}_{32}$), we can gain the allowed range of $r$, which will give a primary restriction on the parameter space of $\{ {\rm Re}\,\tau,{\rm Im}\,\tau,g,\phi^{}_g\}$. Then we continue to diagonalize the neutrino mass matrix $\widetilde{M}^{}_{\nu}$ via $U^{\dag} \widetilde{M}^{}_{\nu} \widetilde{M}^{\dag}_{\nu} U= {\rm Diag}\left\{m^{2}_{1}, m^{2}_{2}, m^{2}_{3}\right\}$, and obtain the lepton mixing matrix $U$ at the electroweak scale. In the standard parametrization~\cite{Tanabashi:2018oca}, we have
	\begin{eqnarray}
	U = \left(
	\begin{matrix}
	c^{}_{12}c^{}_{13} && s^{}_{12}c^{}_{13} && s^{}_{13}e^{-{\rm i}\delta}_{} \\
	-s^{}_{12}c^{}_{23}-c^{}_{12}s^{}_{23}s^{}_{13}e^{{\rm i}\delta} && c^{}_{12}c^{}_{23}-s^{}_{12}s^{}_{23}s^{}_{13}e^{{\rm i}\delta}_{} && s^{}_{23}c^{}_{13} \\
	s^{}_{12}s^{}_{23}-c^{}_{12}c^{}_{23}s^{}_{13}e^{{\rm i}\delta} && -c^{}_{12}s^{}_{23}-s^{}_{12}c^{}_{23}s^{}_{13}e^{{\rm i}\delta} &&
	c^{}_{23}c^{}_{13}
	\end{matrix}\right)\; ,
	\label{eq:UPMNS}
	\end{eqnarray}
	where $c^{}_{ij} \equiv \cos \theta^{}_{ij}$ and $s^{}_{ij} \equiv \sin \theta^{}_{ij}$ (for $ij =12,13,23$) have been defined. Comparing the obtained values of $s^{2}_{ij}$ with their individual $3\sigma$ ($1\sigma$) ranges from global-fit results, we can find out the $3\sigma$ ($1\sigma$) allowed parameter space in our model.

\begin{table}[t]
	\begin{center}
		\vspace{-0.25cm} \caption{The best-fit values, the
			1$\sigma$ and 3$\sigma$ intervals, together with the values of $\sigma^{}_{i}$ being the symmetrized $1\sigma$ uncertainties, for three neutrino mixing angles $\{\theta^{}_{12}, \theta^{}_{13}, \theta^{}_{23}\}$, two mass-squared differences $\{\Delta m^2_{21}, \Delta m^2_{31}~{\rm or}~\Delta m^2_{32}\}$ and the Dirac CP-violating phase $\delta$ from a global-fit analysis of current experimental data~\cite{Esteban:2018azc}.} \vspace{0.5cm}
		\begin{tabular}{c|c|c|c|c}
			\hline
			\hline
			Parameter & Best fit & 1$\sigma$ range &  3$\sigma$ range & $\sigma^{}_{i}$ \\
			\hline
			\multicolumn{5}{c}{Normal neutrino mass ordering
				$(m^{}_1 < m^{}_2 < m^{}_3)$} \\ \hline
			$\sin^2_{}\theta^{}_{12}$
			& $0.310$ & 0.298 --- 0.323 &  0.275 --- 0.350 & 0.0125 \\
			$\sin^2_{}\theta^{}_{13}$
			& $0.02241$ & 0.02176 --- 0.02307 &  0.02046 --- 0.02440  & 0.000655 \\
			$\sin^2_{}\theta^{}_{23}$
			& $0.558$  & 0.525 --- 0.578 &  0.427 --- 0.609  & 0.0265 \\
			$\delta~[^\circ]$ &  $222$ & 194 --- 260 &  141 --- 370 & 33 \\
			$\Delta m^2_{21}~[10^{-5}~{\rm eV}^2]$ &  $7.39$ & 7.19 --- 7.60 & 6.79 --- 8.01 & 0.205 \\
			$\Delta m^2_{31}~ [10^{-3}~{\rm eV}^2]$ &  $+2.523$ & +2.493 --- +2.555 & +2.432 --- +2.618 & 0.031 \\\hline
			\multicolumn{5}{c}{Inverted neutrino mass ordering
				$(m^{}_3 < m^{}_1 < m^{}_2)$} \\ \hline
			$\sin^2_{}\theta^{}_{12}$
			& $0.310$ & 0.298 --- 0.323 &  0.275 --- 0.350 & 0.0125\\
			$\sin^2_{}\theta^{}_{13}$
			& $0.02261$ & 0.02197 --- 0.02328 &  0.02066 --- 0.02461 & 0.000655 \\
			$\sin^2_{}\theta^{}_{23}$
			& $0.563$  & 0.537 --- 0.582 &  0.430 --- 0.612 & 0.0225  \\
			$\delta~[^\circ]$ &  $285$ & 259 --- 309 &  205 --- 354 & 25 \\
			$\Delta m^2_{21}~[10^{-5}~{\rm eV}^2]$ &  $7.39$ & 7.19 --- 7.60 & 6.79 --- 8.01 & 0.205 \\
			$\Delta m^2_{32}~[10^{-3}~{\rm eV}^2]$ &  $-2.509$ & $-2.539$ --- $-2.477$  & $-2.603$ --- $-2.416$ & $0.031$ \\ \hline\hline
		\end{tabular}
		\label{table:gfit}
	\end{center}
\end{table}
	
	\item In order to measure how well the model is consistent with current experimental data, we can further construct the $\chi^2$-function, namely,
	\begin{eqnarray}
	\chi^2_{}(p^{}_{i}) = \sum^{}_{j}\left(\frac{q^{}_{j}(p^{}_{i})-q^{\rm bf}_{j}}{\sigma^{}_{j}}\right)^2_{} \; ,
	\label{eq:chi}
	\end{eqnarray}
	where $p^{}_{i} \in \{{\rm Re}\,\tau, {\rm Im}\,\tau, g, \phi^{}_g,v^{}_{\rm u}g^{}_1/\sqrt{2}\}$ stand for the model parameters, $q^{}_j(p^{}_i)$ denote the model predictions for the observables $\{\sin^2\theta^{}_{12}, \sin^2\theta^{}_{13}, \sin^2\theta^{}_{23}, \Delta m^2_{21}, \Delta m^2_{31}(\Delta m^2_{32})\}$ and $q^{\rm bf}_j$ are their best-fit values from the global analysis in Ref.~\cite{Esteban:2018azc}. The uncertainties $\sigma^{}_j$ are derived by symmetrizing $1\sigma$ uncertainties from the global-fit results, which have already been given in Table~\ref{table:gfit}. We can calculate the minimum value of the $\chi^2_{}$-function, which corresponds to the best-fit values of our model parameters. 
\end{itemize}

After implementing the above procedure, we find that ten of the twelve scenarios derived in Sec.~\ref{sec:models} can accommodate the latest experimental data. In the following, we will mainly focus on four typical scenarios and present the detailed numerical results.

\subsection{L1N1}
\begin{figure}[t!]
	\centering		\includegraphics[width=0.98\textwidth]{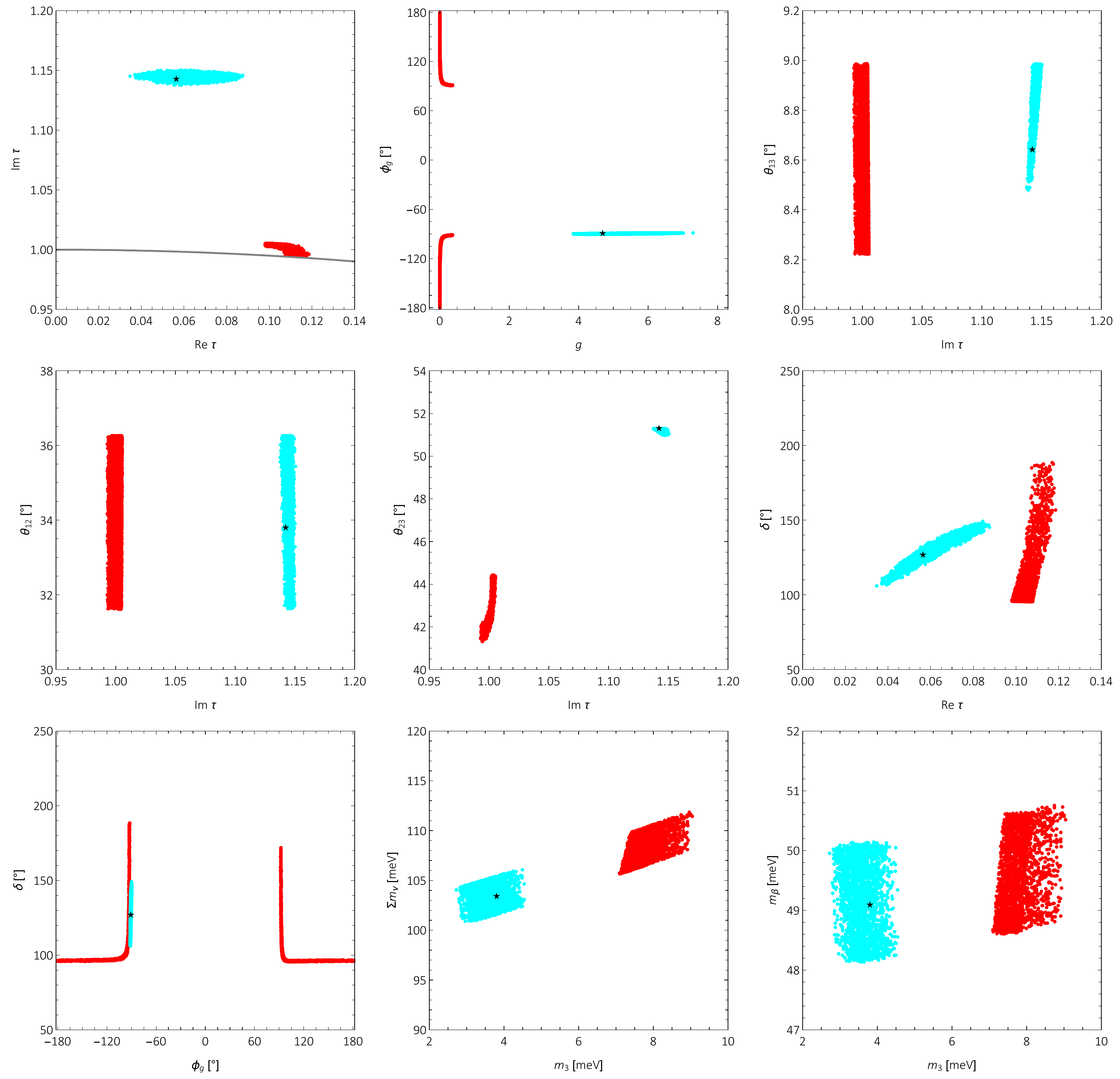}
	\vspace{-0.3cm}
	\caption{Allowed parameter space of the model parameters $\{{\rm Re}\,\tau, {\rm Im}\,\tau\}$ and $\{g, \phi^{}_g\}$ and the constrained ranges of low-energy observables in the IO case in {\bf L1N1}, where the $3\sigma$ ranges of mixing angles and neutrino mass-squared differences from the global-fit analysis of neutrino oscillation data have been input~\cite{Esteban:2018azc}. The red and cyan dots represent two distinct regions. The gray curve in the top-left panel denotes the lower boundary of the fundamental domain ${\cal G}$. The best-fit values from our $\chi^2$-fit analysis are indicated by the black stars.}
	\label{fig:L1N1} 
\end{figure}

The allowed parameter space and the constrained ranges of low-energy observables in {\bf L1N1} are shown in Fig.~\ref{fig:L1N1}, where one can find that  {\bf L1N1} is consistent with experimental data at the $3\sigma$ level in the IO case.

As one can see from Fig.~\ref{fig:L1N1}, the parameter space of {\bf L1N1} is separated into two distinct regions denoted by red and cyan dots, respectively. The red region corresponds to $\alpha^{}_3<\alpha^{}_1<\alpha^{}_2$. In this region, the allowed parameter space of $\{{\rm Re}\,\tau, {\rm Im}\,\tau\}$ is very narrow, i.e., $0.10<{\rm Re}\,\tau <0.12$ and ${\rm Im}\,\tau \sim 1$. The top-middle panel of Fig.~\ref{fig:L1N1} shows that the parameter space of $\{g, \phi^{}_g\}$ is approximately centered on the axis $\phi^{}_g=0$ in the red region, i.e., we can reverse the sign of $\phi^{}_g$ and keep other parameters unchanged and obtain similar predictions for the low-energy observables. Note that this relation only holds for small value of $g$ since the dependence of oscillation parameters on $\phi^{}_g$ is not significant if $g$ is small. While if $g>1$, we could not find such an approximate symmetry any more, which will be seen in some of the following cases. The small value of $g$ in {\bf L1N1} indicates that the first matrix in the square brackets in Eq.~(\ref{eq:MD1}) dominantly contributions to neutrino masses. Especially, the minimum of $g$ can be as small as $7\times10^{-3}_{}$. As a reasonable estimation, we can neglect the second matrix in the square brackets in Eq.~(\ref{eq:MD1}). Then the mass matrix $M^{}_{\nu}$ can be easily diagonalized by the unitary matrix $U^{(0)}_{\nu}$ expressed as
\begin{eqnarray}
 U^{(0)}_{\nu}=\left(
 \begin{matrix}
 0 && 1 && 0 \\
 \dfrac{\sqrt{2}}{2} && 0 && \dfrac{\sqrt{2}}{2} \\
\dfrac{\sqrt{2}}{2} && 0 && -\dfrac{\sqrt{2}}{2} 
 \end{matrix}\right) \; ,
 \label{eq:approUnu1}
\end{eqnarray}
and three neutrino mass eigenvalues turn out to be 
\begin{eqnarray}
m^{}_{1} &=& \frac{\sqrt{2}}{4}v^{}_{\rm u}g^{}_1|\sqrt{3}Y^{}_2-Y^{}_1| \; , \label{eq:approeigen1}\\
m^{}_{2} &=& \frac{\sqrt{2}}{2}v^{}_{\rm u}g^{}_1|Y^{}_{1}|\; , \label{eq:approeigen2}\\
m^{}_{3} &=& \frac{\sqrt{2}}{4}v^{}_{\rm u}g^{}_1|\sqrt{3}Y^{}_2+Y^{}_1| \; .\label{eq:approeigen3}
\end{eqnarray} 
We can select some specific values of $\{{\rm Re}\,\tau, {\rm Im}\,\tau\}$ in the allowed parameter space and substitute them into the above equations. For example, if ${\rm Re}\,\tau=0.103$ and $ {\rm Im}\,\tau =1.003$, we will obtain $r \equiv \Delta m^2_{21}/|\Delta m^2_{32}| \approx 0.0252$. As a comparison, the exact value of $r$ without approximation is 0.0272. Therefore, we can give a good description of neutrino masses by only considering the first matrix in Eq.~(\ref{eq:MD1}). However, if $g$ is exactly zero, it will not lead to realistic mixing angles. Let us also take ${\rm Re}\,\tau=0.103$ and $ {\rm Im}\,\tau =1.003$ for instance. For simplicity, here we do not consider the RG running effects. The unitary matrix $U^{}_{l}$ in the charged lepton sector in this case is
\begin{eqnarray}
U^{}_l = \left(
\begin{matrix}
0.697 && -0.498-0.130\,{\rm i} && -0.493+0.079\,{\rm i} \\
0.490+0.069\,{\rm i} && -0.140-0.076\,{\rm i} && 0.854 \\
0.499-0.141\,{\rm i} && 0.843 && -0.137+0.046\,{\rm i} \\
\end{matrix}\right) \; ,
\label{eq:ulnum}
\end{eqnarray}
Then by using Eqs.~(\ref{eq:approUnu1}) and (\ref{eq:ulnum}), we can obtain the mixing matrix $U=U^{\dag}_{l}U^{(0)}_{\nu}$, where one can read $\sin^2_{}\theta^{}_{12} = 0.497$, which has already exceeded the upper bound of its $3\sigma$ allowed range. Therefore, we need to consider the higher order corrections from $\widetilde{g}$. Since $\phi^{}_g \sim \pm 180^{\circ}_{}$ when $g$ is extremely small, we can neglect the phase of $\widetilde{g}$ and simply assume $\widetilde{g} \approx -g$. Up to the first order of $g$, we have
\begin{eqnarray}
M^{}_{\nu}M^{\dag}_{\nu} \approx \left(\begin{matrix}
1.488 && -(0.293-0.152\,{\rm i})g && -(1.591+0.295\,{\rm i})g \\
-(0.293+0.152\,{\rm i})g && 0.742 && 0.708-0.314\,{\rm i}g \\
-(1.591-0.295\,{\rm i})g && 0.708+0.314\,{\rm i}g && 0.742 \\
\end{matrix}\right) \; .
\label{eq:o1Mnu}
\end{eqnarray}
The unitary matrix $U^{(1)}_{\nu}$ up to the first order of $g$ can be written as
\begin{eqnarray}
U^{(1)}_{\nu} = U^{(0)}_{\nu}+g \Delta U^{}_{\nu} \; ,
\label{eq:o1U1}
\end{eqnarray}
where $U^{(0)}_{\nu}$ takes the form in Eq.~(\ref{eq:approUnu1}) and $\Delta U^{}_{\nu}$ is assumed to be
\begin{eqnarray}
\Delta U^{}_{\nu} \approx \left(\begin{matrix}
x^{}_{11} + {\rm i}\,y^{}_{11} && x^{}_{12} + {\rm i}\,y^{}_{12}&& x^{}_{13} + {\rm i}\,y^{}_{13} \\
x^{}_{21} + {\rm i}\,y^{}_{21} && x^{}_{22} + {\rm i}\,y^{}_{22} && x^{}_{23} + {\rm i}\,y^{}_{23} \\
x^{}_{31} + {\rm i}\,y^{}_{31} && x^{}_{32} +{\rm i}\,y^{}_{32} && x^{}_{33} + {\rm i}\,y^{}_{33} \\
\end{matrix}\right) \; ,
\label{eq:deltau1}
\end{eqnarray}
where $x^{}_{ij}$ and $y^{}_{ij}$ (for $i,j=1,2,3$) are real parameters. Since each column of $U^{(1)}_{\nu}$ should be the eigenvector of $M^{}_{\nu}M^{\dag}_{\nu}$, $x^{}_{ij}$ and $y^{}_{ij}$ can be determined by comparing the first order terms of $g$ in both the left and right hand sides of the following equation
\begin{eqnarray}
M^{}_{\nu}M^{\dag}_{\nu}U^{(1)}_{\nu} = U^{(1)}_{\nu} {\rm Diag}\{m^{2}_{1},m^{2}_{2},m^{2}_{3}\} \; ,
\label{eq:eigenequation}
\end{eqnarray}
where $m^{}_{i}$ (for $i=1,2,3$) are the zeroth-order neutrino mass eigenvalues shown in Eqs.~(\ref{eq:approeigen1})-(\ref{eq:approeigen3}). After $x^{}_{ij}$ and $y^{}_{ij}$ are obtained, $\Delta U^{}_{\nu}$ can be expressed as
\begin{eqnarray}
\Delta U^{}_{\nu} \approx \left(\begin{matrix}
36.22+2.753\,{\rm i} &&      0                   && -0.631-0.217\,{\rm i} \\
-0.472-0.156\,{\rm i} &&   -25.21+1.797\,{\rm i} && 0.472-0.156\,{\rm i} \\
0.472+0.156\,{\rm i} &&    -26.10+2.104\,{\rm i} && 0.472-0.156\,{\rm i} \\
\end{matrix}\right) \; .
\label{eq:deltau1value}
\end{eqnarray} 
Then by using Eqs.~(\ref{eq:approUnu1}), (\ref{eq:ulnum}), (\ref{eq:o1U1}) and (\ref{eq:deltau1value}), we finally obtain the mixing matrix $U=U^{\dag}_{l}U^{(1)}_{\nu}$, where one can get the relations between $\sin^{2}_{}\theta^{}_{ij}$ (for $ij=13,12,23$) and $g$ as
\begin{eqnarray}
\sin^{2}_{}\theta^{}_{13} &\approx& 0.0220(1+3.641g+3.428g^2_{}) \; ,  \nonumber \\
\sin^{2}_{}\theta^{}_{12} &\approx& 0.497(1-73.20g+1336g^2_{}) \; ,   \nonumber  \\
\sin^{2}_{}\theta^{}_{23} &\approx& 0.496(1-1.189g+0.897g^2_{}) \; ,
\label{eq:mixg}
\end{eqnarray}
where one can clearly see that the large coefficients in front of $g$ and $g^2_{}$ in the second equation lead to significant corrections to $\sin^2_{}\theta^{}_{12}$. If $g=6.71 \times 10^{-3}_{}$, Eq.~(\ref{eq:mixg}) gives $\sin^{2}_{}\theta^{}_{13}\approx0.0221$, $\sin^{2}_{}\theta^{}_{12}\approx0.283$ and $\sin^{2}_{}\theta^{}_{12}\approx0.490$, which are in good agreement with accurate numerical results.

As can be clearly seen in Fig.~\ref{fig:L1N1}, the predicted values of $\theta^{}_{23}$ in the red region are located in the first octant, varying from $41^{\circ}_{}$ to $44.5^{\circ}_{}$, while the values of $\theta^{}_{12}$ and $\theta^{}_{13}$ can cover all the $3\sigma$ allowed ranges from global-fit results. There are two sources that determine the value of $\delta$ in our model, which are ${\rm Re}\,\tau$ and $\phi^{}_g$. We have also presented the correlations between $\delta$ and these two parameters in Fig.~\ref{fig:L1N1}. It is shown that the allowed range of $\delta$ is $90^{\circ}_{}<\delta<190^{\circ}_{}$ and the value of $\delta$ is not much affected by the sign of $\phi^{}_g$. On the other hand, the sum of neutrino masses $\sum m^{}_{\nu} \equiv m^{}_{1}+m^{}_{2}+m^{}_{3}$ is about $110~{\rm meV}$, as can be seen in the bottom-middle panel of Fig.~\ref{fig:L1N1}. In addition, with the neutrino mass spectrum and mixing parameters known, we can predict the effective neutrino mass for beta decays, i.e., 
\begin{eqnarray}
m^{}_\beta \equiv \sqrt{m^2_1 |U^{}_{e1}|^2 + m^2_2 |U^{}_{e2}|^2 + m^2_3 |U^{}_{e3}|^2} \; .
\label{eq:mbeta}
\end{eqnarray} 
The bottom-right panel in Fig.~\ref{fig:L1N1} shows that $m^{}_{\beta}$ is around $50~{\rm meV}$ in the red region of {\bf L1N1}. Currently the most stringent restriction on $m^{}_{\beta}$ comes from the KATRIN experiment, which indicates $m^{}_\beta < 1.1~{\rm eV}$ at the $90\%$ confidence level~\cite{Aker:2019uuj, Aker:2019qfn}. With more data accumulated in KATRIN, the upper bound will be improved to $m^{}_\beta < 0.2~{\rm eV}$. However, it is still far away from the value of $m^{}_{\beta}$ predicted in {\bf L1N1}.

The cyan region in Fig.~\ref{fig:L1N1} corresponds to another hierarchy of $\alpha^{}_1$, $\alpha^{}_2$ and $\alpha^{}_3$, which is $\alpha^{}_3<\alpha^{}_2<\alpha^{}_1$. In this region, the parameter space of $\{{\rm Re}\,\tau, {\rm Im}\,\tau\}$ is restricted to $0.03<{\rm Re}\,\tau<0.09$ and ${\rm Im}\,\tau \sim 1.15$. The value of $g$ in the cyan region is larger than one, to be specific, $3.7<g<7.5$, indicating that the contributions to the neutrino masses and flavor mixing from the second matrix in Eq.~(\ref{eq:MD1}) become dominate. The allowed value of $\phi^{}_g$ is tightly constrained, which is $\phi^{}_g \sim 90^{\circ}_{}$. Unlike the red region, we could not find the symmetric parameter space of $\phi^{}_g$ near $-90^{\circ}_{}$ due to the large value of $g$. The predicted value of $\theta^{}_{23}$ is very close to the upper bound of its $3\sigma$ range, i.e., $\theta^{}_{23} \sim 51^{\circ}_{}$. Furthermore, the sum of three neutrino masses $\sum m^{}_{\nu}$ and the effective neutrino mass for beta decays $m^{}_{\beta}$ are slightly smaller than those predicted in the red region. 

Implementing the $\chi^2_{}$-fit analysis, we obtain the minimum $\chi^{2}_{\rm min} =  3.23$ of the $\chi^2_{}$-function, corresponding to the following best-fit values of model parameters
\begin{eqnarray}
{\rm Re}\,\tau = 0.0565\;, \quad {\rm Im}\,\tau = 1.142 \;, \quad g = 4.697 \;, \quad \phi^{}_{g} =89.7^{\circ}_{} \; , \quad v^{}_{\rm u}g^{}_1/\sqrt{2} = 9.61~{\rm meV} \; .
\label{eq:paraL1N1}
\end{eqnarray}
Note that here we use $v^{}_{\rm u}g^{}_1/\sqrt{2}$ to denote the absolute scale of neutrino masses at the electroweak scale after the RG running. The best-fit values above, together with the values of $m^{}_e$, $m^{}_{\mu}$ and $m^{}_{\tau}$ lead to $v^{}_{\rm d} \alpha^{}_1 /\sqrt{2} = 1.70~{\rm GeV}$, $\alpha^{}_2/\alpha^{}_1 = 3.98\times 10^{-2}_{}$ and $\alpha^{}_3/\alpha^{}_1 = 1.43\times 10^{-4}$. With these best-fit values of model parameters, we get the neutrino mass spectrum $m^{}_1 = 49.51~{\rm meV}$, $m^{}_2 = 50.15~{\rm meV}$ and $m^{}_3 = 3.80~{\rm meV}$, three mixing angles $\theta^{}_{12} = 33.79^{\circ}$, $\theta^{}_{13} = 8.64^{\circ}$ and $\theta^{}_{23} =51.29^{\circ}$, and the Dirac CP-violating phase $\delta = 127^{\circ}$. Furthermore, the best-fit value of $m^{}_{\beta}$ is $49.09~{\rm meV}$. 

An interesting feature of {\bf L1N1} is that the value of $\alpha^{}_3$ is much smaller than $\alpha^{}_1$ and $\alpha^{}_2$, indicating that the third column of $M^{}_l$ in Eq.~(\ref{eq:Me1}), which is generated by $Y^{(4)}_{{\bf 3}^{\prime}_{}}$, is highly suppressed when compared with the other two columns. Therefore, if we replace $Y^{(4)}_{{\bf 3}^{\prime}_{}}$ with $Y^{(6)}_{\bf 3}$ or $Y^{(6)}_{{\bf 3}^{\prime}_{},2}$ and keep the remaining parts unchanged, i.e., we change the model {\bf L1N1} to {\bf L2N1} or {\bf L3N1}, the numerical results will almost be the same as those obtained in {\bf L1N1}. For illustration, we still use the best-fit values of $\{{\rm Re}\,\tau, {\rm Im}\,\tau,g,\phi^{}_g,v^{}_{\rm u}g^{}_1/\sqrt{2}\}$ shown in Eq.~(\ref{eq:paraL1N1}) to calculate the predictions for mixing angles and the Dirac CP-violating phase in {\bf L2N1} (The neutrino masses should keep invariant since we do not adjust the mass matrix in the neutrino sector). The results are
\begin{eqnarray}
\theta^{}_{12} = 33.79^{\circ}\;, \quad \theta^{}_{13} = 8.64^{\circ} \;, \quad \theta^{}_{23} =51.29^{\circ} \;, \quad  \delta = 127^{\circ}_{} \;,
\label{eq:obser2N1}
\end{eqnarray}
where one could hardly see any difference between them and the corresponding best-fit values in {\bf L1N1}. As a result, we get similar parameter space and constrained ranges of observables in {\bf L2N1} and {\bf L3N1}, which can also fit the experimental data in the IO case at the $3\sigma$ level. 

\begin{figure}[t!]
	\centering		\includegraphics[width=0.98\textwidth]{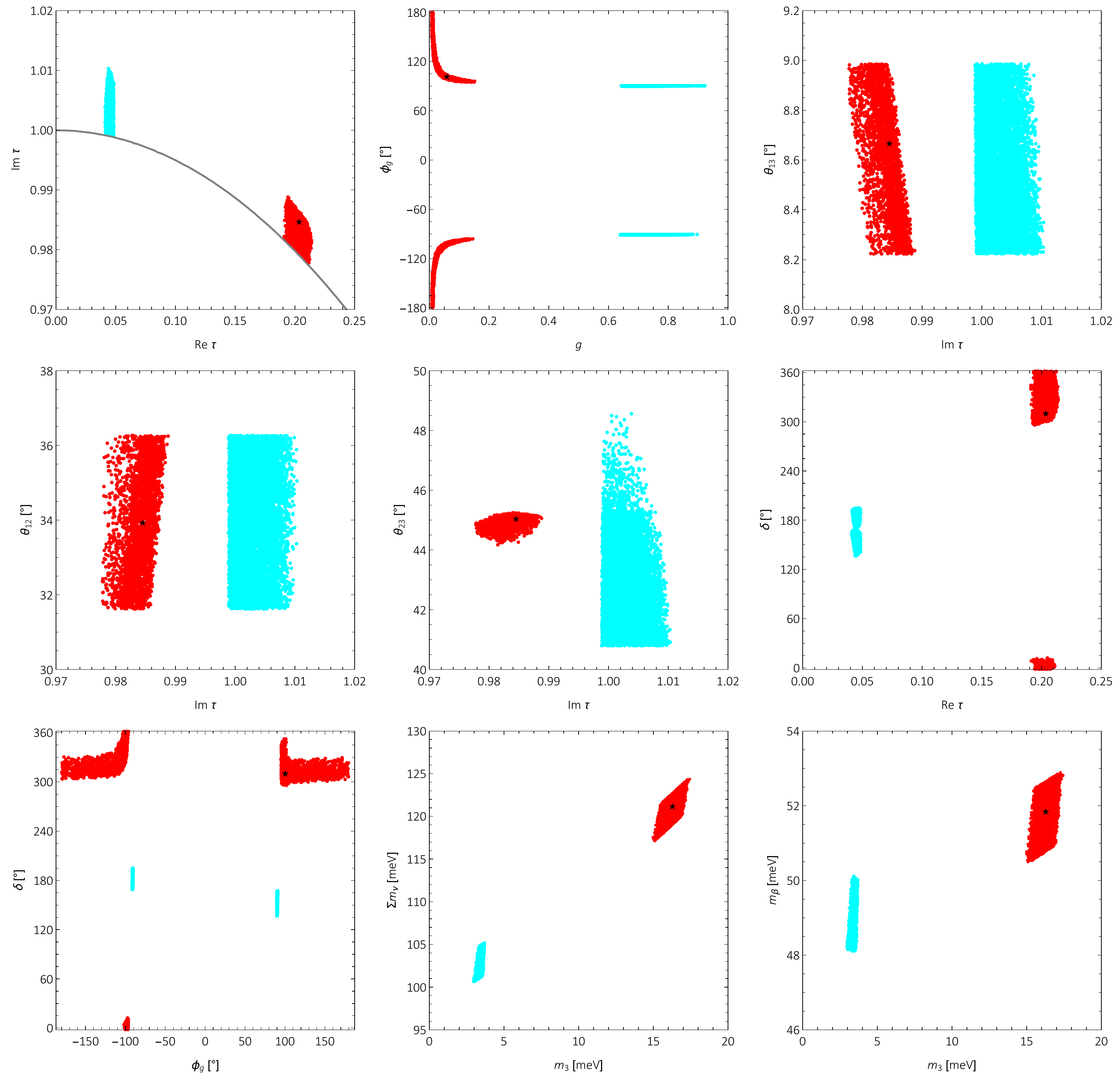}
	\vspace{-0.3cm}
	\caption{Allowed parameter space of the model parameters $\{{\rm Re}\,\tau, {\rm Im}\,\tau\}$ and $\{g, \phi^{}_g\}$ and the constrained ranges of low-energy observables in the IO case in {\bf L3N1}, where the $3\sigma$ ranges of mixing angles and neutrino mass-squared differences from the global-fit analysis of neutrino oscillation data have been input~\cite{Esteban:2018azc}. The red and cyan dots represent two distinct regions. The gray curve in the top-left panel denotes the lower boundary of the fundamental domain ${\cal G}$. The best-fit values from our $\chi^2$-fit analysis are indicated by the black stars.}
	\label{fig:L3N1} 
\end{figure}

\subsection{{\bf L3N1}}
As we have discussed in the last subsection, {\bf L3N1} can have similar parameter space to {\bf L1N1}. However, this is not the unique parameter space of {\bf L3N1}. In Fig.~\ref{fig:L3N1}, we show another two distinct regions in the allowed parameter space of {\bf L3N1}, which are still compatible with experiments in the IO case at the $3\sigma$ level. The red region in Fig.~\ref{fig:L3N1} corresponds to the hierarchy $\alpha^{}_1<\alpha^{}_3<\alpha^{}_2$, where ${\rm Re}\,\tau \sim 0.2$ and ${\rm Im}\,\tau \sim 0.985$. The value of $g$ is a small number, $g<0.2$, and there are also two approximate symmetric parts in the parameter space of $\{g, \phi^{}_g\}$. The value of $\theta^{}_{23}$ is constrained in the range where $44.3^{\circ}_{} < \theta^{}_{23} <45.3^{\circ}_{}$, which is smaller than the lower bound of its $1\sigma$ range. Most of the predicted values of $\delta$ are located in the fourth octant, from $300^{\circ}_{}$ to $360^{\circ}_{}$, which is preferred by the global-fit results of $\delta$~\cite{Esteban:2018azc}. The sum of neutrino masses $\sum m^{}_{\nu} \sim 121~{\rm meV}$ is very close to the upper bound $\sum m^{}_{\nu}<120~{\rm meV}$ from Planck observations~\cite{Aghanim:2018eyx}, thus can be easily tested in future experiments. Another region denoted by the cyan color corresponds to the hierarchy $\alpha^{}_2<\alpha^{}_1<\alpha^{}_3$. In this region, we have ${\rm Re}\,\tau \sim 0.05$ and $1 < {\rm Im}\,\tau< 1.01$. The parameter space of $\{g,\phi^{}_g\}$ is still almost centered on the axis $\phi^{}_g=0$, with $0.64<g<0.93$ and $\phi^{}_g  \approx \pm 90^{\circ}_{}$. Different from the red region, $\theta^{}_{23}$ in the cyan region varies in a large range from $40.8^{\circ}_{}$ to $48.6^{\circ}_{}$, and the value of $\delta$ is located in the region where $135^{\circ}_{}<\delta<195^{\circ}_{}$. Since now $\alpha^{}_{2}$ is the smallest parameter among $\alpha^{}_1$, $\alpha^{}_2$ and $\alpha^{}_3$, we can change the modular forms $f^{}_{\mu}(\tau)$ into $Y^{(4)}_{{\bf 3}^{\prime}_{}}$ or $Y^{(6)}_{\bf 3}$. Correspondingly we will arrive at the model {\bf L5N1} or {\bf L6N1} which can still compatible with the experimental data at the $3\sigma$ level in the IO case, just like the red region in {\bf L3N1}.

Based on the $\chi^2_{}$-analysis, we obtain the following best-fit values of parameters in {\bf L3N1}
\begin{eqnarray}
{\rm Re}\,\tau = 0.204\;, \quad {\rm Im}\,\tau = 0.985 \;, \quad g = 0.0592 \;, \quad \phi^{}_{g} =101^{\circ}_{} \;, \quad v^{}_{\rm u}g^{}_1/\sqrt{2} = 44.0~{\rm meV} \; ,
\label{eq:paraL3N1}
\end{eqnarray}
which corresponds to the minimum $\chi^{2}_{\rm min}=5.31$. Then the best-fit values for other parameters are $v^{}_{\rm d} \alpha^{}_1 /\sqrt{2} = 0.717 ~{\rm MeV}$, $\alpha^{}_2/\alpha^{}_1 = 1.12\times 10^{3}_{}$ and $\alpha^{}_3/\alpha^{}_1 = 82.1$. With these best-fit values, we get the neutrino mass spectrum $m^{}_1 = 52.06~{\rm meV}$, $m^{}_2 = 52.77~{\rm meV}$ and $m^{}_3 = 16.29~{\rm meV}$, three mixing angles $\theta^{}_{12} = 33.93^{\circ}$, $\theta^{}_{13} = 8.66^{\circ}$ and $\theta^{}_{23} =45.03^{\circ}$, the Dirac CP-violating phase $\delta = 308^{\circ}$ and the effective neutrino mass for beta decays $m^{}_{\beta}=51.82~{\rm meV}$.

\begin{figure}[t!]
	\centering		\includegraphics[width=0.98\textwidth]{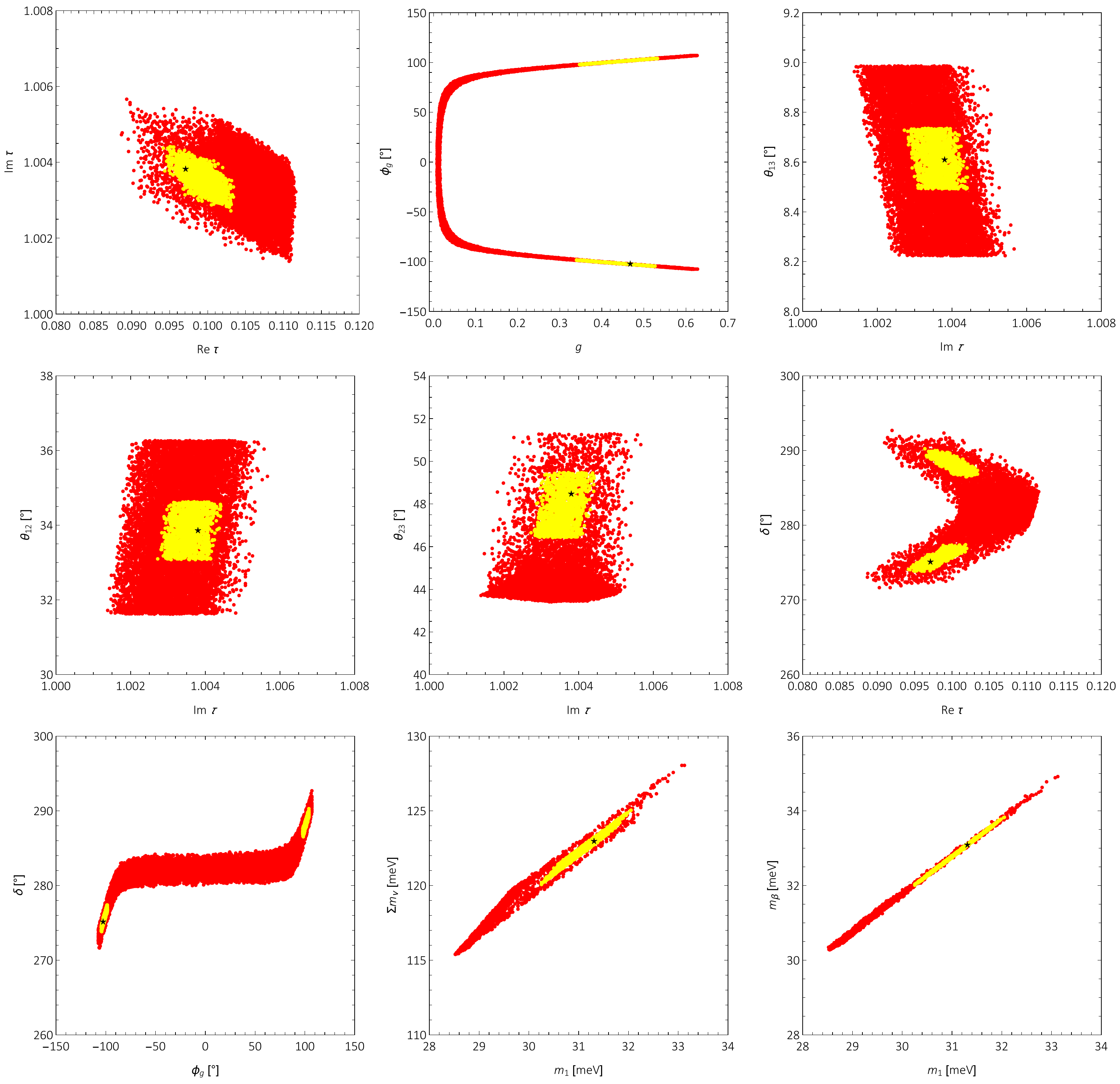}
	\vspace{-0.3cm}
	\caption{Allowed parameter space of the model parameters $\{{\rm Re}\,\tau, {\rm Im}\,\tau\}$ and $\{g, \phi^{}_g\}$ and the constrained ranges of low-energy observables in the NO case in {\bf L1N2}, where the $3\sigma$ (red dots) and $1\sigma$ (yellow dots) ranges of mixing angles and neutrino mass-squared differences from the global-fit analysis of neutrino oscillation data have been input~\cite{Esteban:2018azc}. The best-fit values from our $\chi^2$-fit analysis are indicated by the black stars.}
	\label{fig:L1N2} 
\end{figure}

\subsection{L1N2}
The allowed parameter space and the constrained ranges of low-energy observables in {\bf L1N2} are shown in Fig.~\ref{fig:L1N2}, where one can find that {\bf L1N2} is consistent with experimental data at the $1\sigma$ level in the NO case. It is very interesting that the $3\sigma$ allowed parameter space of ${\rm Re}\,\tau$ and  ${\rm Im}\,\tau$ is analogous to the one in {\bf L1N1}. Since the charged-lepton mass matrices in {\bf L1N1} and {\bf L1N2} take the same form, we conclude that these two models lead to similar $U^{}_l$ in the charged-lepton sector, and the different mixing pattern of these two models arises only from the distinct forms of $M^{}_{\nu}$. The value of $g$ can go from $8.6 \times 10^{-3}_{}$ to $0.64$ within the $3\sigma$ range. While at the $1\sigma$ level, there are two distinct regions in the parameter space of $\{g, \phi^{}_g\}$ which are approximately centered on the axis $\phi^{}_g=0$. The value of $\phi^{}_g$ is around $\pm 100^{\circ}_{}$, thus can have important contributions to the CP violation. As can be seen in the bottom-left panel of Fig.~\ref{fig:L1N2}, the two distinct regions of $\phi^{}_g$ correspond to two different ranges of $\delta$ which are $[273^{\circ}_{},278^{\circ}_{}]$ and $[286^{\circ}_{},291^{\circ}_{}]$, respectively. 

We find that {\bf L1N2} fits the experimental data very well with the minimum $\chi^{2}_{\rm min} = 0.0533$, corresponding to
\begin{eqnarray}
{\rm Re}\,\tau = 0.0971\;, \quad {\rm Im}\,\tau = 1.004 \;, \quad g = 0.4679 \;, \quad \phi^{}_{g} = -102^{\circ}_{} \;, \quad v^{}_{\rm u}g^{}_1/\sqrt{2} = 39.1~{\rm meV} \; ,
\label{eq:paraL1N2}
\end{eqnarray}
which together with the values of $m^{}_e$, $m^{}_{\mu}$ and $m^{}_{\tau}$ lead to $v^{}_{\rm d} \alpha^{}_1 /\sqrt{2} = 9.28\times 10^{-2}_{}~{\rm GeV}$, $\alpha^{}_2/\alpha^{}_1 = 9.07$ and $\alpha^{}_3/\alpha^{}_1 = 2.15\times 10^{-3}$. In addition, we get the neutrino mass spectrum $m^{}_1 = 31.31~{\rm meV}$, $m^{}_2 = 32.46~{\rm meV}$ and $m^{}_3 = 59.18~{\rm meV}$, three mixing angles $\theta^{}_{12} = 33.85^{\circ}$, $\theta^{}_{13} = 8.61^{\circ}$ and $\theta^{}_{23} =48.45^{\circ}$, and the Dirac CP-violating phase $\delta = 275^{\circ}$. The effective neutrino mass for beta decays turns out to be $33.08~{\rm meV}$.

Note that $\alpha^{}_{3} \ll \alpha^{}_1,\alpha^{}_2$ is also satisfied in {\bf L1N2}, which means we can still replace $Y^{(4)}_{\bf 3}$ with $Y^{(6)}_{\bf 3}$ or $Y^{(6)}_{{\bf 3}^{\prime}_{},2}$ for $f^{}_{\tau}(\tau)$ and change {\bf L1N2} into {\bf L2N2} or {\bf L3N2}. As a result, {\bf L1N2}, {\bf L2N2} and {\bf L3N2} have analogous parameter space and are all consistent with the neutrino oscillation experimental data within the $1\sigma$ range in the NO case.

\subsection{L3N2}
\begin{figure}[t!]
	\centering		\includegraphics[width=0.98\textwidth]{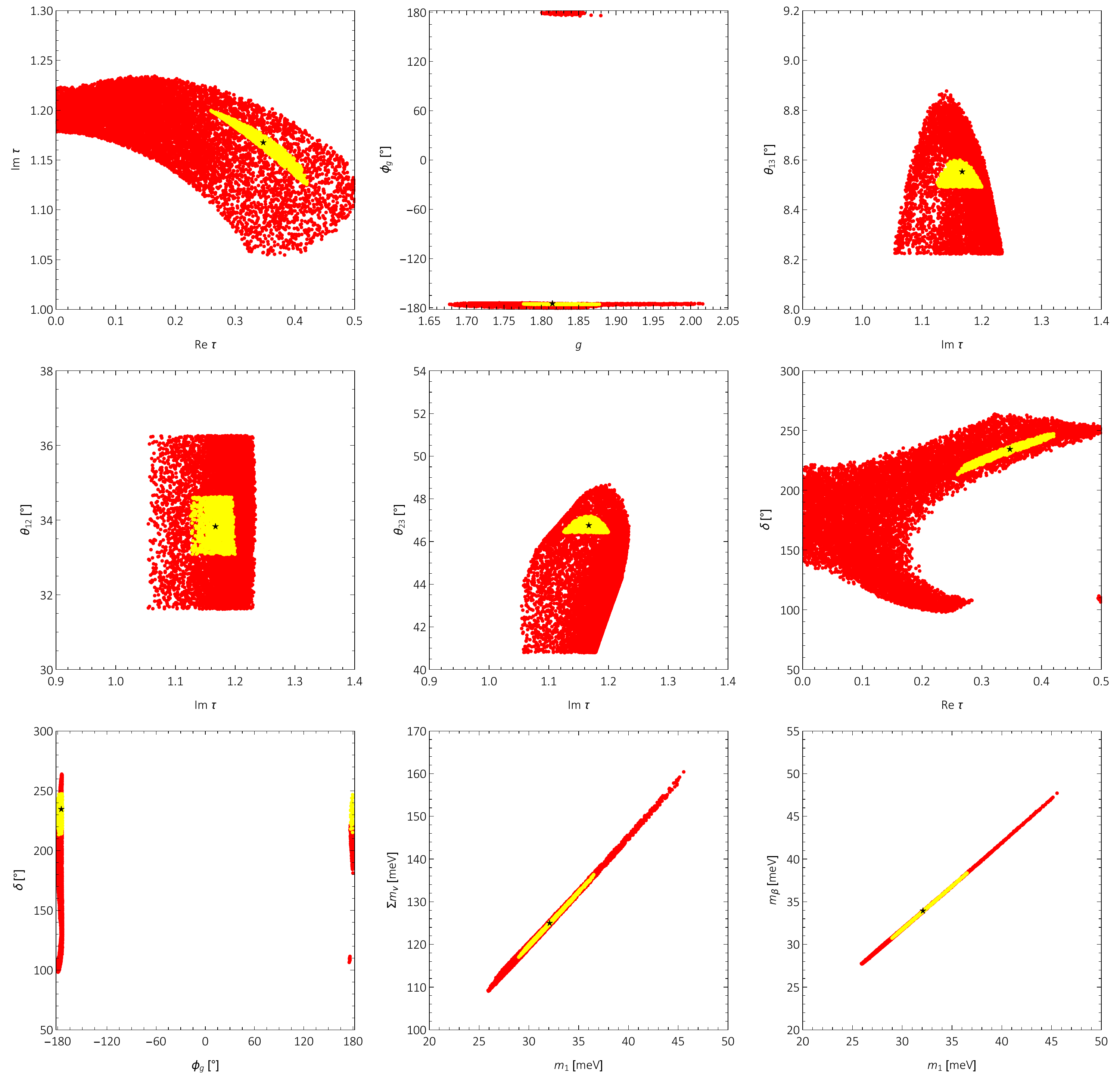}
	\caption{Allowed parameter space of the model parameters $\{{\rm Re}\,\tau, {\rm Im}\,\tau\}$ and $\{g, \phi^{}_g\}$ and the constrained ranges of low-energy observables in the NO case in {\bf L3N2}, where the $3\sigma$ (red dots) and $1\sigma$ (yellow dots) ranges of mixing angles and neutrino mass-squared differences from the global-fit analysis of neutrino oscillation data have been input~\cite{Esteban:2018azc}. The best-fit values from our $\chi^2$-fit analysis are indicated by the black stars.}
	\label{fig:L3N2} 
	\vspace{0.3cm}
\end{figure}
Apart from the region which is similar to that in {\bf L1N2}, there is another distinct region in the allowed parameter space of {\bf L3N2}, which corresponds to the hierarchy $\alpha^{}_2<\alpha^{}_3<\alpha^{}_1$. As can be observed from the top-left panels of Fig.~\ref{fig:L3N2}, the whole range $[0,0.5]$ of ${\rm Re}\,\tau$ is allowed at the $3\sigma$ level and ${\rm Im}\,\tau$ varies from 1.05 to 1.24. $\phi^{}_g$ is close to $\pm 180^{\circ}_{}$, which can hardly generate large CP violation in the neutrino sector. Therefore the main source of the CP violation in {\bf L3N2} is ${\rm Re\,\tau}$. The values of $\theta^{}_{13}$ and $\theta^{}_{23}$ can reach the lower bounds of their respective $3\sigma$ and $1\sigma$ ranges, but are not able to touch the upper bounds. The sum of neutrino masses $\sum m^{}_{\nu}$ predicted in this allowed region is relatively large, with a maximum $(\sum m^{}_{\nu})^{}_{\rm max}=162~{\rm meV}$. After calculating the $\chi^2_{}$-function, we find that the minimum $\chi^{2}_{\rm min} = 1.33$ is obtained with the following best-fit values of the model parameters
\begin{eqnarray}
{\rm Re}\,\tau = 0.347\;, \quad {\rm Im}\,\tau = 1.167 \;, \quad g = 1.815 \;, \quad \phi^{}_{g} = -175^{\circ}_{} \;, \quad v^{}_{\rm u}g^{}_1/\sqrt{2} = 24.2~{\rm meV} \; ,
\label{eq:paraL3N2}
\end{eqnarray}
which together with the values of $m^{}_e$, $m^{}_{\mu}$ and $m^{}_{\tau}$ lead to $v^{}_{\rm d} \alpha^{}_1 /\sqrt{2} = 1.73~{\rm GeV}$, $\alpha^{}_2/\alpha^{}_1 = 1.96 \times 10^{-4}_{}$ and $\alpha^{}_3/\alpha^{}_1 = 6.26\times 10^{-2}$. We can also get the neutrino masses $m^{}_1 = 32.10~{\rm meV}$, $m^{}_2 = 33.24~{\rm meV}$ and $m^{}_3 = 59.57~{\rm meV}$, three mixing angles $\theta^{}_{12} = 33.82^{\circ}$, $\theta^{}_{13} = 8.55^{\circ}$ and $\theta^{}_{23} =46.74^{\circ}$, the Dirac CP-violating phase $\delta = 234^{\circ}$ and the effective neutrino mass for beta decays $m^{}_{\beta}= 33.88~{\rm meV}$. Furthermore, since now $\alpha^{}_{2} \ll \alpha^{}_1,\alpha^{}_3$, we can replace $f^{}_{\mu}(\tau) \sim Y^{(4)}_{\bf 3}$ with $Y^{(6)}_{\bf 3}$ or $Y^{(6)}_{{\bf 3}^{\prime}_{},2}$ in the charged-lepton sector, and change {\bf L3N2} to {\bf L5N2} or {\bf L6N2}. Then we conclude that {\bf L5N2} and {\bf L6N2} are also consistent with the experiments at the $1\sigma$ level in the NO case. 

\begin{figure}
	\centering		\includegraphics[width=0.96\textwidth]{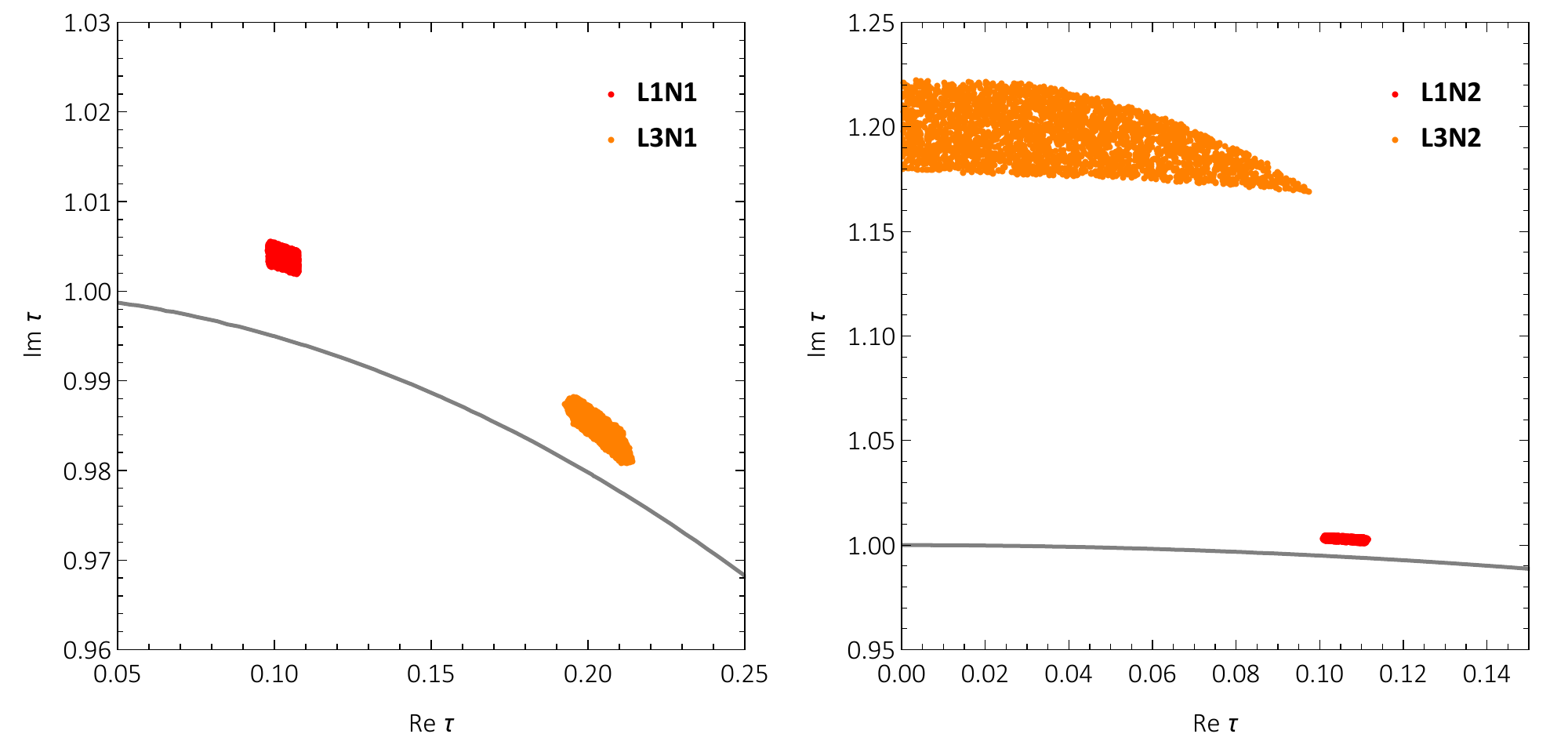}
	\vspace{-0.3cm}
	\caption{The $3\sigma$ allowed parameter space of the model parameters $\{{\rm Re}\,\tau, {\rm Im}\,\tau\}$ with $\widetilde{g}$ being a real number in {\bf L1N1}, {\bf L3N1}, {\bf L1N2} and {\bf L3N2}, where the red and yellow dots denote the models {\bf L1N1} ({\bf L1N2}) and {\bf L3N1} ({\bf L3N2}) in the left (right) panel, respectively. The gray curves represent the lower boundary of the fundamental domain ${\cal G}$.}
	\label{fig:figParagcp} 
\end{figure}

As a summary, we study the low-energy phenomenology of the twelve Dirac neutrino mass models derived in Sec.~\ref{sec:models}. We find that five of them are consistent with experimental data in the IO case at the $3\sigma$ level, which are {\bf L1N1}, {\bf L2N1}, {\bf L3N1}, {\bf L5N1} and {\bf L6N1}. While five models can fit the experiments in the NO case at the $1\sigma$ level, which are {\bf L1N2}, {\bf L2N2}, {\bf L3N2}, {\bf L5N2} and {\bf L6N2}. 

\renewcommand\arraystretch{1.1}
\begin{table}[t]
	\begin{center}
		\vspace{-0.25cm} \caption{The best-fit values of free model parameters and observables in {\bf L1N1}, {\bf L3N1}, {\bf L1N2} and {\bf L3N2} with the gCP symmetry.} \vspace{0.5cm}
		\begin{tabular}{c|c|c|c|c|c}
			\hline
			\hline
			\multicolumn{2}{c|}{} & {\bf L1N1} & {\bf L3N1} & {\bf L1N2} & {\bf L3N2} \\
			\hline
			\multirow{7}*{\rotatebox[]{90}{Free model parameters}} 
			& ${\rm Re}\,\tau$ & 0.103 & 0.204 & 0.106 & $1.61\times10^{-3}_{}$ \\
			~ & ${\rm Im}\,\tau$ & 1.004 & 0.985 & 1.003 & 1.205 \\
			~ & $\widetilde{g}$              & $-7.19\times10^{-3}_{}$  & $-0.0111$ & 0.0114 & $-1.820$ \\ 
			~ & $v^{}_{\rm d}\alpha^{}_{1}/\sqrt{2}$ $[{\rm GeV}]$ &  $9.27\times10^{-2}_{}$ & $7.27 \times 10^{-3}_{}$ & $9.27\times10^{-2}_{}$ & 1.78 \\
			~ & $\alpha^{}_2/\alpha^{}_1$ & 9.08 & $1.11\times10^3_{}$ & 9.07 & $2.83\times10^{-4}_{}$ \\
			~ & $\alpha^{}_3/\alpha^{}_1$ & $2.16\times10^{-3}_{}$ & 81.0 & $2.17\times 10^{-3}_{}$ & $5.07\times10^{-2}_{}$ \\ 
			~ & $v^2_{\rm u}g^2_1/(2\Lambda)$ $[{\rm meV}]$ & 41.57 & 43.84 & 41.64 & 23.26 \\
			\hline
			\multirow{9}*{\rotatebox[]{90}{Observables}} 
			& $m^{}_1$ $[{\rm meV}]$& 49.96 & 51.82 & 29.17 & 28.78 \\
			~ & $m^{}_2$ $[{\rm meV}]$& 50.70 & 52.52 & 30.42 & 30.04 \\
			~ & $m^{}_3$ $[{\rm meV}]$& 7.62 & 16.18 & 58.00 & 57.89 \\
			~ & $\theta^{}_{12}$ $[{}^{\circ}_{}]$ & 33.99 & 33.65 & 34.43 & 34.33 \\
			~ & $\theta^{}_{13}$ $[{}^{\circ}_{}]$ & 8.62 & 8.61 & 8.61 & 8.36 \\
			~ & $\theta^{}_{23}$ $[{}^{\circ}_{}]$& 44.33 & 45.13 & 43.83 & 42.94 \\ 
			~ & $\delta$ $[{}^{\circ}_{}]$& 96.4 & 313 & 282 & 181 \\
			~ & $m^{}_{\beta}$ $[{\rm meV}]$& 49.64 & 51.51 & 30.50 & 30.08 \\
			\hline
			& $\chi^2_{\rm min}$  & 7.595 & 5.328 & 8.889 & 16.00 \\
			\hline
			\multicolumn{2}{c|}{Mass ordering}   & IO  &  IO & NO & NO \\
			\hline
			\hline
		\end{tabular}
		\label{table:gcp}
	\end{center}
\end{table}
\renewcommand\arraystretch{1}

\begin{table}[t]
	\begin{center}
		\vspace{-0.25cm} \caption{Comparison between the best-fit values of $\{r, \theta^{}_{12}, \theta^{}_{13}, \theta^{}_{23}, \delta\}$ at the electroweak scale $m^{}_Z = 91.2~{\rm GeV}$ after taking into account the RG running effects and their corresponding values at the GUT scale $\Lambda^{}_{\rm GUT} = 2\times 10^{16}_{}~{\rm GeV}$  in {\bf L1N1}, {\bf L3N1}, {\bf L1N2} and {\bf L3N2}, where $\tan\beta=10$ and $m^{}_{\rm SUSY}=10~{\rm TeV}$ are assumed.} \vspace{0.5cm}
		\begin{tabular}{c|c|c|c|c|c|c}
			\hline
			\hline
			\multicolumn{2}{c|}{} & $r$ & $\theta^{}_{12}/^\circ$ & $\theta^{}_{13}/^\circ$ & $\theta^{}_{23}/^\circ$ & $\delta/^\circ$  \\
			\hline
			\multirow{2}*{{\bf L1N1}}
			& $m^{}_Z$ & 0.0294 & 33.79 & 8.64 & 51.29 & 127\\
			~ & $\Lambda^{}_{\rm GUT}$ & 0.0302 & 31.23 & 8.65 & 51.35 & 125 \\
			\hline
			\multirow{2}*{{\bf L3N1}}
			& $m^{}_Z$ & 0.0292 & 33.93 & 8.66 & 45.03 & 308\\
			~ & $\Lambda^{}_{\rm GUT}$ & 0.0307 & 31.94 & 8.68 & 45.11 & 309 \\
			\hline
			\multirow{2}*{{\bf L1N2}}
			& $m^{}_Z$ & 0.0291 & 33.85 & 8.61 & 48.45 & 275 \\
			~ & $\Lambda^{}_{\rm GUT}$& 0.0295 & 32.94 & 8.59 & 48.33 & 276 \\
			\hline
			\multirow{2}*{{\bf L3N2}}
			& $m^{}_Z$ & 0.0296 & 33.82 & 8.55 & 46.74 & 234 \\
			~ & $\Lambda^{}_{\rm GUT}$ & 0.0298 & 32.84 & 8.54 & 46.62 & 235 \\
			\hline
			\hline
		\end{tabular}
		\label{table:RGE}
	\end{center}
\end{table}
\subsection{Models with the gCP symmetry}
In this subsection, we discuss the cases where $\widetilde{g}$ is real, i.e., $\phi^{}_g=0$ or $180^\circ_{}$ due to the gCP symmetry. As can be clearly seen from Figs.~\ref{fig:L1N1}-\ref{fig:L3N2}, there exist some regions where $\phi^{}_g=0$ or $180^\circ_{}$ in the $3\sigma$ allowed parameter space of all the models {\bf L1N1}, {\bf L1N2}, {\bf L3N1} and {\bf L3N2} discussed above. It is then reasonable to expect that these models are still consistent with the global-fit results even if the gCP symmetry is taken into consideration. Therefore, we assume $\widetilde{g}$ to be real and seek for the feasible parameter space again in {\bf L1N1}, {\bf L1N2}, {\bf L3N1} and {\bf L3N2} following the similar numerical strategy shown in Sec.~\ref{sec:lowphe}. 

The numerical results of the $3\sigma$ allowed parameter space of $\{{\rm Re}\,\tau,{\rm Im}\,\tau\}$ in {\bf L1N1}, {\bf L1N2}, {\bf L3N1} and {\bf L3N2} are shown in Fig.~\ref{fig:figParagcp}, where we can find that the allowed parameter space of ${\rm Re}\,\tau$ and ${\rm Im}\,\tau$ of all these four models shrinks compared with that without gCP, which conforms our expectation since the inclusion of gCP can only make the complex parameters in the model real. On the other hand, the allowed parameter space of {\bf L1N1} and {\bf L1N2} departures from the lower boundary of the fundamental domain. It has been proved that the CP-conserving values of $\tau$ are the imaginary axis and the boundary of the fundamental domain in Ref.~\cite{Novichkov:2019sqv}, therefore {\bf L1N1} and {\bf L1N2} will predict nonzero values of $\delta$. In addition, the $3\sigma$ allowed parameter space of ${\rm Re}\,\tau$ in {\bf L3N2} is no longer the whole range of $[0,0.5]$. The maximal value of ${\rm Re}\,\tau$  in {\bf L3N2} with gCP symmetry turns out to be $0.10$. In Table~\ref{table:gcp}, we present the best-fit values of free parameters as well as low-energy observables in {\bf L1N1}, {\bf L3N1}, {\bf L1N2} and {\bf L3N2} with the gCP symmetry, where we can find that all these four models can be compatible with the global-fit results only at the $3\sigma$ level due to the further constraints from the gCP symmetry.

\subsection{Corrections from RG running effects}
Before closing up this section, let us discuss to what extend the RG running effects can modify our model. For illustration, we also calculate the predictions for $\{r, \theta^{}_{12}, \theta^{}_{13}, \theta^{}_{23}, \delta\}$ at the GUT scale, i.e., without considering the RG running effects, by using the best-fit values of free model parameters we have obtained in Eqs. (\ref{eq:paraL1N1}), (\ref{eq:paraL3N1}), (\ref{eq:paraL1N2}) and (\ref{eq:paraL3N2}), and compare them with their respective best-fit values at the electroweak scale. The results are shown in Table~\ref{table:RGE}, where one can observe that the implication of RG running effects to the IO case ({\bf L1N1} and {\bf L3N1}) is more significant than that to the NO case ({\bf L1N2} and {\bf L3N2}). This can be understood considering two relatively large neutrino masses in the IO case are nearly-degenerate, which will enhance the radiative corrections. In addition, $\theta^{}_{12}$ is most sensitive to the RG running effects among all the observables. The value of $\theta^{}_{12}$ obtained at $m^{}_{Z}$ is about $2^{\circ}_{}$ ($1^{\circ}_{}$) larger than that obtained at $\Lambda^{}_{\rm GUT}$ in the IO (NO) case.

\begin{figure}
	\centering		\includegraphics[width=0.96\textwidth]{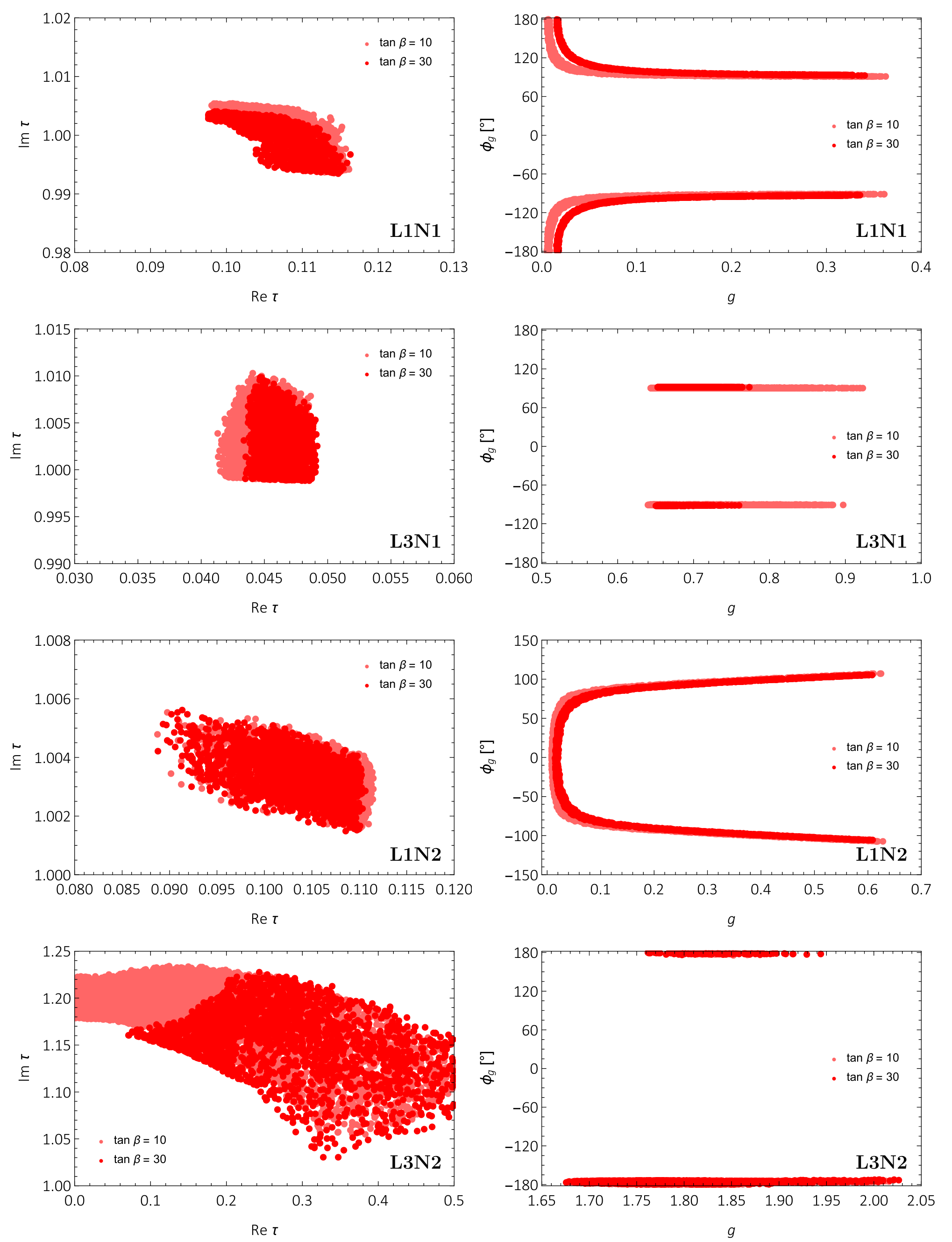}
	\vspace{-0.3cm}
	\caption{The $3\sigma$ allowed parameter space of the model parameters $\{{\rm Re}\,\tau, {\rm Im}\,\tau\}$ and $\{g, \phi^{}_g\}$ in {\bf L1N1}, {\bf L3N1}, {\bf L1N2} and {\bf L3N2} under the assumption that $\tan\beta=30$ and $\tan\beta=10$, denoted by dark and light red points, respectively. Note that in the plots of the first two rows, we only retain the red regions in the parameter space of {\bf L1N1} and {\bf L3N1} when $\tan\beta=10$ for convenience of comparison.}
	\label{fig:figParaRGE} 
\end{figure}

On the other hand, corrections from RG running effects could become more significant if we assume a larger value of $\tan\beta$. In order to make this point clearer, we have further found out the $3\sigma$ allowed parameter space of $\{{\rm Re}\,\tau, {\rm Im}\,\tau\}$ and $\{g, \phi^{}_g\}$ in our models when $\tan\beta=30$. The results are presented in Fig.~\ref{fig:figParaRGE}, together with the parameter space with $\tan\beta=10$ we have obtained in previous subsections. One can find that the four typical scenarios {\bf L1N1}, {\bf L3N1}, {\bf L1N2} and {\bf L3N2} we have discussed before are still compatible with experimental data within the $3\sigma$ level even if $\tan\beta=30$, but the allowed parameter space has been reduced. For example, from the first two plots in Fig.~\ref{fig:figParaRGE}, one can find that there remains only one piece of region in the parameter space of {\bf L1N1} when $\tan\beta=30$, while the cyan region in Fig.~\ref{fig:L1N1} is not allowed any more. This is because the predicted value of $\sin^{2}_{}\theta^{}_{12}$ in the cyan region is larger than 0.36, which has already exceeded its $3\sigma$ allowed range. This is also the case for {\bf L3N1}, where the red region in Fig.~\ref{fig:L3N1} is excluded if $\tan\beta=30$.


\begin{figure}[t!]
	\centering		\includegraphics[width=0.98\textwidth]{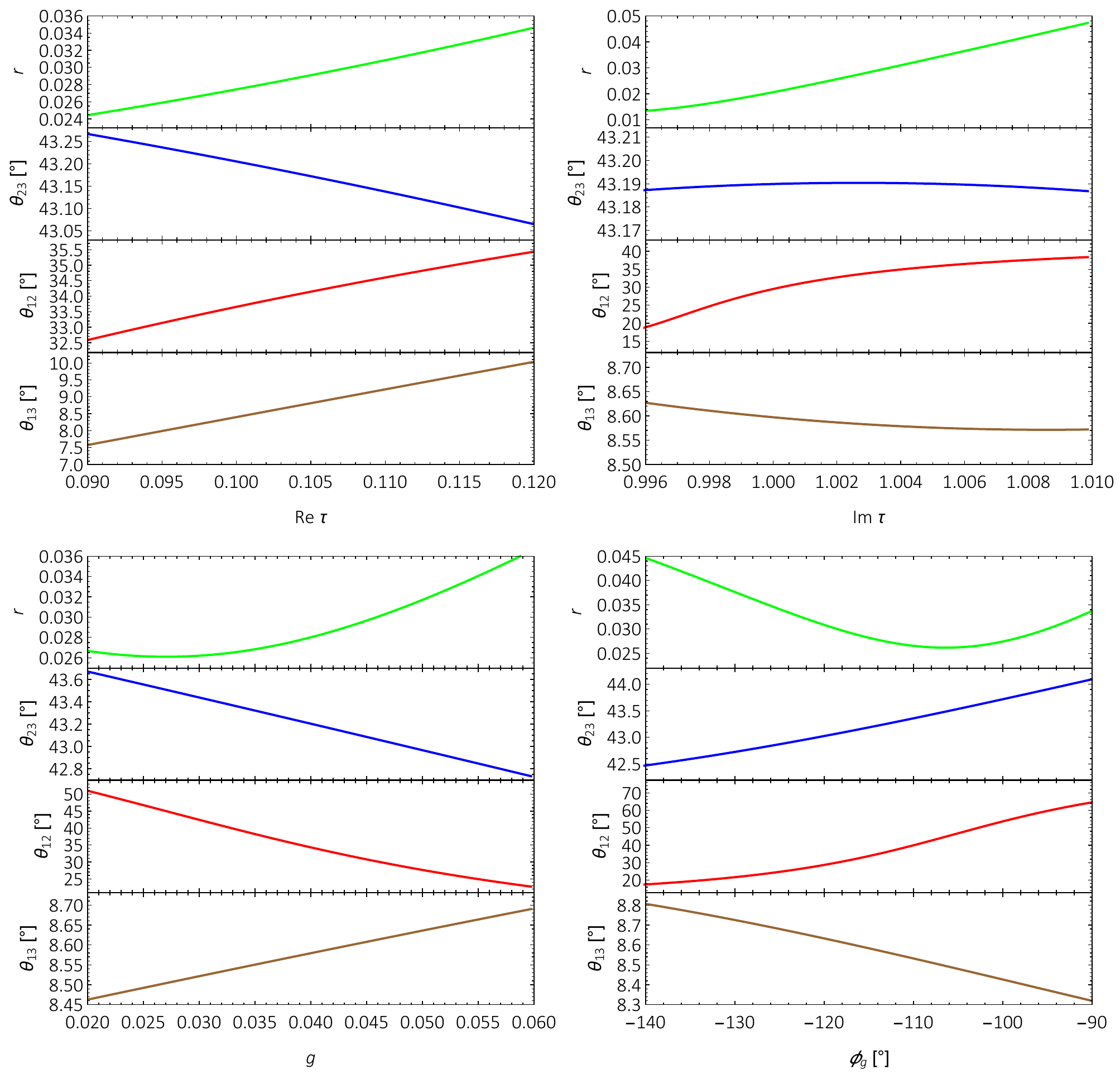}
	\vspace{0cm}
	\caption{The evolution behavior of  $\{r,\theta^{}_{12},\theta^{}_{13},\theta^{}_{23}\}$ against $\{{\rm Re}\,\tau, {\rm Im}\,\tau, g, \phi^{}_g\}$ in {\bf L1N1} with $\tan\beta=30$. For each plot, we require only one parameter to change and keep other parameters staying at their respective best-fit values shown in Eq.~(\ref{eq:bftan30}). }
	\label{fig:angle} 
\end{figure}

However, in some cases the allowed parameter space with $\tan\beta=30$ only slightly deviates from that with $\tan\beta=10$, which seems a little bit confusing considering the large modifications to $\theta^{}_{12}$ from the RG running effects. To illustrate this issue, we study the evolution behavior of the ratio $r$ and three mixing angles $\{\theta^{}_{12},\theta^{}_{13},\theta^{}_{23}\}$ towards the model parameters $\{{\rm Re}\,\tau, {\rm Im}\,\tau, g, \phi^{}_g\}$. Let us also take {\bf L1N1} as an example. In Fig.~\ref{fig:angle} we show the evolution behavior of  $\{r,\theta^{}_{12},\theta^{}_{13},\theta^{}_{23}\}$ against $\{{\rm Re}\,\tau, {\rm Im}\,\tau, g, \phi^{}_g\}$ near their respective best-fit values which are
\begin{eqnarray}
{\rm Re}\,\tau = 0.102\;, \quad {\rm Im}\,\tau = 1.003 \;, \quad g = 0.404 \;, \quad \phi^{}_{g} = -111^{\circ}_{} \;.
\label{eq:bftan30}
\end{eqnarray}
In order to clearly show the relation between $\{r,\theta^{}_{12},\theta^{}_{13},\theta^{}_{23}\}$ and each individual parameter, for each plot we keep only one parameter varying and all the other parameters staying at their respective best-fit values. As can be seen in Fig.~\ref{fig:angle}, $\theta^{}_{13}$ and $\theta^{}_{23}$ are not sensitive to all the four parameters, while the value of $\theta^{}_{12}$ can significantly decrease as $g$ goes larger or ${\rm Im}\,\tau$ and $\phi^{}_g$ go smaller. Therefore, we can always slightly low down the values of ${\rm Im}\,\tau$ and $\phi^{}_{g}$ in the red region of the parameter space in {\bf L1N1} with $\tan\beta=10$, or rise up the value of $g$, and arrive at the allowed parameter space with $\tan\beta=30$. Consequently, the parameter space for $\{{\rm Re}\,\tau, {\rm Im}\,\tau\}$ should move downwards and that for $\{g, \phi^{}_g\}$ should move rightwards, which is exactly what the first two plots in Fig.~\ref{fig:L1N1} look like.

\section{Summary}\label{sec:summary}
Finite modular symmetries have been widely incorporated into seesaw models to account for the Majorana neutrino masses, lepton flavor mixing and CP violation, whereas a detailed study about the modular invariant Dirac neutrino mass models is still lacking. In this paper, we investigate Dirac neutrino mass models with a modular $S^{}_4$ symmetry. 

The superfields for lepton doublets $\widehat{L}$ are set to be the triplet {\bf 3} under the modular $S^{}_4$ symmetry while the superfields for right-handed neutrinos $\widehat{N}^{\rm C}_{}$ can take either {\bf 3} or ${\bf 3}^{\prime}_{}$. In order to forbid the Majorana mass term, we require the weight of $\widehat{N}^{\rm C}_{}$ to be a positive integer. There is some freedom in the selection of weights and representations of the modular forms. In the charged-lepton sector, we set $f^{}_{e}(\tau)$ to be $Y^{(2)}_{{\bf 3}^{\prime}_{}}$ while the forms of $f^{}_{\mu}(\tau)$ and $f^{}_{\tau}(\tau)$ are selected from $Y^{(4)}_{\bf 3}$, $Y^{(4)}_{{\bf 3}^{\prime}_{}}$, $Y^{(6)}_{\bf 3}$ and $Y^{(6)}_{{\bf 3}^{\prime}_{},2}$. In the neutrino sector, we only adopt the modular forms with the lowest non-trivial weight two, since the modular forms with higher weights will bring more free parameters into our model.  As a result, we obtain six different charged-lepton mass matrices and two neutrino mass matrices. Their combinations lead to twelve classes of distinct models with eight real parameters, labeled by {\bf L1N1} --- {\bf L6N2}. As we know, the modular symmetry intrinsically works at some high-energy scales, whereas the values of oscillation parameters are measured at the electroweak scale. Therefore, in order to obtain more accurate predictions for these low-energy observables, we have also considered corrections from the RG running effects in our model. We argue that these effects can not be neglected especially for high energy scales or large values of $\tan\beta$. After the numerical calculation, we find that ten of the twelve models are permitted by current neutrino oscillation experiments. To be specific, {\bf L1N1}, {\bf L2N1}, {\bf L3N1}, {\bf L5N1} and {\bf L6N1} are consistent with experimental data in the IO case at the $3\sigma$ level, while {\bf L1N2}, {\bf L2N2}, {\bf L3N2}, {\bf L5N2} and {\bf L6N2} can fit the experiments in the NO case at the $1\sigma$ level. We get the allowed parameter space of $\tau$ and $\widetilde{g}$, the constrained ranges of three mixing angles $\{\theta^{}_{12},\theta^{}_{13},\theta^{}_{23}\}$, and the predictions for the Dirac CP-violating phase $\delta$, the sum of neutrino masses $\sum m^{}_{\nu}$ and the effective neutrino mass for beta decays $m^{}_{\beta}$ in our model. In addition, we also take the gCP symmetry into consideration, which requires all the coupling constants in our model to be real and reduces the number of free model parameters to seven. We find that all the ten allowed models without gCP symmetry are still consistent with the experimental data at the $3\sigma$ level even if the gCP symmetry is included.

We notice that some of these models can possess similar parameter space and constrained ranges of observables, e.g., {\bf L2N1} and {\bf L3N1} share almost common parameter space with {\bf L1N1}. This is due to the fact that these models only differ in the third column of their charged-lepton mass matrices, whose contributions to flavor mixing are highly suppressed by the small value of $\alpha^{}_3$. Such a relation can also be found in the models like $ \{ {\bf L3N1},{\bf L5N1} ,{\bf L6N1}\}$, $\{{\bf L1N2},{\bf L2N2},{\bf L3N2}\}$ and $ \{ {\bf L3N2},{\bf L5N2} ,{\bf L6N2}\}$.

Since whether neutrinos are Dirac or Majorana particles is not yet determined, it is still worthwhile to investigate the origin of Dirac neutrino masses and flavor mixing, to which the modular symmetries may provide an attractive solution. Other finite modular groups apart from $\Gamma^{}_4 \simeq S^{}_4$ can also be used to construct Dirac neutrino mass models. In addition, it is interesting to discuss whether the modular invariance can give a natural explanation of tiny neutrino Yukawa couplings. We hope to come back to these issues in future works.

\section*{Acknowledgements}
 
I am greatly indebted to Prof.~Shun Zhou for carefully reading this manuscript and useful suggestions. I would also like to thank Dr.~Biswajit Karmakar for helpful discussions. This work was supported in part by the National Natural Science Foundation of China under grant No.~11775232 and No.~11835013.

\newpage
\appendix

\section{The $\Gamma^{}_{4} \simeq S^{}_{4}$ symmetry group}\label{sec:appA}

The permutation symmetry group $S^{}_{4}$ has twenty-four elements and five irreducible representations, which are denoted as ${\bf 1}$, ${\bf 1^{\prime}_{}}$, ${\bf 2}$, ${\bf 3}$ and ${\bf 3^{\prime}_{}}$. In the present work, we choose the same basis for the representation matrices of two generators $S$ and $T$ as in Ref.~\cite{Novichkov:2019sqv}, namely,
\begin{eqnarray}
\begin{array}{cclcl}
{\bf 1} &: ~\quad & \rho(S) = + 1 \; , & ~\quad & \rho(T) = + 1 \; , \\
{\bf 1^{\prime_{}}} &: ~\quad & \rho(S)=-1 \; , & ~\quad & \rho(T)=-1 \; , \\
{\bf 2} &: ~\quad & \displaystyle \rho(S)=\frac{1}{2}\left(\begin{matrix}
-1 &&& \sqrt{3} \\ \sqrt{3} &&& 1
\end{matrix}\right) \; , &~\quad & \rho(T)=\left(\begin{matrix}
1 &&& 0 \\ 0 &&& -1
\end{matrix}\right) \; , \\
{\bf 3} &: ~\quad & \displaystyle \rho(S)=-\frac{1}{2}\left(\begin{matrix}
0 &&& \sqrt{2} &&& \sqrt{2} \\
\sqrt{2} &&& -1 &&& 1 \\
\sqrt{2} &&& 1 &&& -1
\end{matrix}\right) \; , &~\quad & \rho(T)=-\left(\begin{matrix}
1 &&& 0 &&& 0 \\
0 &&& \rm{i} &&& 0 \\
0 &&& 0 &&& \rm{-i}
\end{matrix}\right) \; , \\
{\bf 3}^\prime &: ~\quad & \displaystyle \rho(S)=+\frac{1}{2}\left(\begin{matrix}
0 &&& \sqrt{2} &&& \sqrt{2} \\
\sqrt{2} &&& -1 &&& 1 \\
\sqrt{2} &&& 1 &&& -1
\end{matrix}\right) \; , &~\quad & \rho(T) = + \left(\begin{matrix}
1 &&& 0 &&& 0 \\
0 &&& \rm{i} &&& 0 \\
0 &&& 0 &&& \rm{-i}
\end{matrix}\right) \; .
\end{array}
\end{eqnarray}
In this basis, we can explicitly write down the decomposition rules of the Kronecker products of any two $S^{}_{4}$ multiplets.
\begin{itemize}
	\item For the Kronecker products of the singlet ${\bf 1}$ or ${\bf 1^{\prime}_{}}$ and another $S^{}_{4}$ multiplet ${\bf r} = \{{\bf 1}, {\bf 1^{\prime}_{}}, {\bf 2}, {\bf 3}, {\bf 3}^{\prime}_{}\}$:
	\begin{align}
	{\bf 1}\otimes {\bf r} &= {\bf r} \; ,  \label{eq:1r}\\
	{\bf 1^{\prime}_{}} \otimes {\bf 1^{\prime}_{}} &= {\bf 1} \; ,  \label{eq:1p1p}\\
	(\zeta)^{}_{\bf 1^{\prime}_{}} \otimes
	\left(\begin{matrix}
	\xi^{}_{1} \\ \xi^{}_{2}
	\end{matrix}\right)^{}_{\bf 2}
	&= \left(\begin{matrix}
	\zeta\xi^{}_{2} \\ -\zeta\xi^{}_{1}
	\end{matrix}\right)^{}_{\bf 2}  \; , \label{eq:1p2}\\
	(\zeta)^{}_{\bf 1^{\prime}_{}} \otimes
	\left( \begin{matrix}
	\xi^{}_{1} \\ \xi^{}_{2} \\ \xi^{}_{3}
	\end{matrix}\right)^{}_{\bf 3} &=
	\left( \begin{matrix}
	\zeta\xi^{}_{1} \\ \zeta\xi^{}_{2} \\ \zeta\xi^{}_{3}
	\end{matrix}\right)^{}_{\bf 3^{\prime}_{}} \; ,  \label{eq:1p3}\\
	(\zeta)^{}_{\bf 1^{\prime}_{}} \otimes
	\left( \begin{matrix}
	\xi^{}_{1} \\ \xi^{}_{2} \\ \xi^{}_{3}
	\end{matrix}\right)^{}_{\bf 3^{\prime}_{}} &=
	\left( \begin{matrix}
	\zeta\xi^{}_{1} \\ \zeta\xi^{}_{2} \\ \zeta\xi^{}_{3}
	\end{matrix}\right)^{}_{\bf 3} \; ; \label{eq:1p3p}
	\end{align}
	\item For the Kronecker products of the doublet {\bf 2} and another $S^{}_{4}$ multiplet:
	\begin{align}
	\left(\begin{matrix}
	\zeta^{}_{1} \\ \zeta^{}_{2}
	\end{matrix}\right)^{}_{\bf 2} \otimes
	\left(\begin{matrix}
	\xi^{}_{1} \\ \xi^{}_{2}
	\end{matrix}\right)^{}_{\bf 2} &= (\zeta^{}_{1}\xi^{}_{1}+\zeta^{}_{2}\xi^{}_{2})^{}_{\bf 1} \oplus (\zeta^{}_{1}\xi^{}_{2}-\zeta^{}_{2}\xi^{}_{1})^{}_{\bf 1^{\prime}_{}} \oplus \left(\begin{matrix}
	\zeta^{}_{2}\xi^{}_{2}-\zeta^{}_{1}\xi^{}_{1} \\
	\zeta^{}_{1}\xi^{}_{2}+\zeta^{}_{2}\xi^{}_{1}
	\end{matrix}\right)^{}_{\bf 2} \; , \label{eq:22} \\
	\left(\begin{matrix}
	\zeta^{}_{1} \\ \zeta^{}_{2}
	\end{matrix}\right)^{}_{\bf 2} \otimes
	\left(\begin{matrix}
	\xi^{}_{1} \\ \xi^{}_{2} \\ \xi^{}_{3}
	\end{matrix}\right)^{}_{\bf 3} &=
	\left(\begin{matrix}
	\zeta^{}_{1}\xi^{}_{1} \\ (\sqrt{3}/2)\zeta^{}_{2}\xi^{}_{3}-(1/2)\zeta^{}_{1}\xi^{}_{2} \\ (\sqrt{3}/2)\zeta^{}_{2}\xi^{}_{2}-(1/2)\zeta^{}_{1}\xi^{}_{3}
	\end{matrix}\right)^{}_{\bf 3} \oplus	\left(\begin{matrix}
	-\zeta^{}_{2}\xi^{}_{1} \\ (\sqrt{3}/2)\zeta^{}_{1}\xi^{}_{3}+(1/2)\zeta^{}_{2}\xi^{}_{2} \\ (\sqrt{3}/2)\zeta^{}_{1}\xi^{}_{2}+(1/2)\zeta^{}_{2}\xi^{}_{3}
	\end{matrix}\right)^{}_{\bf 3^{\prime}_{}} \; , \label{eq:23} \\
	\left(\begin{matrix}
	\zeta^{}_{1} \\ \zeta^{}_{2}
	\end{matrix}\right)^{}_{\bf 2} \otimes
	\left(\begin{matrix}
	\xi^{}_{1} \\ \xi^{}_{2} \\ \xi^{}_{3}
	\end{matrix}\right)^{}_{\bf 3^{\prime}_{}} &=
	\left(\begin{matrix}
	-\zeta^{}_{2}\xi^{}_{1} \\ (\sqrt{3}/2)\zeta^{}_{1}\xi^{}_{3}+(1/2)\zeta^{}_{2}\xi^{}_{2} \\ (\sqrt{3}/2)\zeta^{}_{1}\xi^{}_{2}+(1/2)\zeta^{}_{2}\xi^{}_{3}
	\end{matrix}\right)^{}_{\bf 3} \oplus	\left(\begin{matrix}
	\zeta^{}_{1}\xi^{}_{1} \\ (\sqrt{3}/2)\zeta^{}_{2}\xi^{}_{3}-(1/2)\zeta^{}_{1}\xi^{}_{2} \\ (\sqrt{3}/2)\zeta^{}_{2}\xi^{}_{2}-(1/2)\zeta^{}_{1}\xi^{}_{3}
	\end{matrix}\right)^{}_{\bf 3^{\prime}_{}} \; ; \label{eq:23p}
	\end{align}
	\item For the Kronecker products of the triplet ${\bf 3}$ or ${\bf 3^{\prime}_{}}$ and another $S^{}_{4}$ triplet:
	\begin{align}
	\left(\begin{matrix}
	\zeta^{}_{1} \\ \zeta^{}_{2} \\ \zeta^{}_{3}
	\end{matrix}\right)^{}_{\bf 3} \otimes
	\left(\begin{matrix}
	\xi^{}_{1} \\ \xi^{}_{2} \\ \xi^{}_{3}
	\end{matrix}\right)^{}_{\bf 3} = &\left(\begin{matrix}
	\zeta^{}_{1} \\ \zeta^{}_{2} \\ \zeta^{}_{3}
	\end{matrix}\right)^{}_{\bf 3^{\prime}_{}} \otimes
	\left(\begin{matrix}
	\xi^{}_{1} \\ \xi^{}_{2} \\ \xi^{}_{3}
	\end{matrix}\right)^{}_{\bf 3^{\prime}_{}} \nonumber \\
	=&~(\zeta^{}_{1}\xi^{}_{1}+\zeta^{}_{2}\xi^{}_{3}+\zeta^{}_{3}\xi^{}_{2})^{}_{\bf 1} \oplus \left( \begin{matrix}
	\zeta^{}_{1}\xi^{}_{1}-(1/2)(\zeta^{}_{2}\xi^{}_{3}+\zeta^{}_{3}\xi^{}_{2}) \\ (\sqrt{3}/2)(\zeta^{}_{2}\xi^{}_{2}+\zeta^{}_{3}\xi^{}_{3})
	\end{matrix}\right)^{}_{\bf 2} \nonumber \\
	& \oplus \left(\begin{matrix}
	\zeta^{}_{3}\xi^{}_{3}-\zeta^{}_{2}\xi^{}_{2} \\
	\zeta^{}_{1}\xi^{}_{3}+\zeta^{}_{3}\xi^{}_{1} \\
	-\zeta^{}_{1}\xi^{}_{2}-\zeta^{}_{2}\xi^{}_{1}
	\end{matrix}\right)^{}_{\bf 3} \oplus \left(\begin{matrix}
	\zeta^{}_{3}\xi^{}_{2}-\zeta^{}_{2}\xi^{}_{3} \\ \zeta^{}_{2}\xi^{}_{1}-\zeta^{}_{1}\xi^{}_{2} \\
	\zeta^{}_{1}\xi^{}_{3}-\zeta^{}_{3}\xi^{}_{1}
	\end{matrix}\right)^{}_{\bf 3^{\prime}_{}} \; ,  \label{eq:33} \\
	\left(\begin{matrix}
	\zeta^{}_{1} \\ \zeta^{}_{2} \\ \zeta^{}_{3}
	\end{matrix}\right)^{}_{\bf 3} \otimes
	\left(\begin{matrix}
	\xi^{}_{1} \\ \xi^{}_{2} \\ \xi^{}_{3}
	\end{matrix}\right)^{}_{\bf 3^{\prime}_{}}
	=&~(\zeta^{}_{1}\xi^{}_{1}+\zeta^{}_{2}\xi^{}_{3}+\zeta^{}_{3}\xi^{}_{2})^{}_{\bf 1^{\prime}_{}} \oplus \left( \begin{matrix}
	(\sqrt{3}/2)(\zeta^{}_{2}\xi^{}_{2}+\zeta^{}_{3}\xi^{}_{3})\\ -\zeta^{}_{1}\xi^{}_{1}+(1/2)(\zeta^{}_{2}\xi^{}_{3}+\zeta^{}_{3}\xi^{}_{2})
	\end{matrix}\right)^{}_{\bf 2} \nonumber \\
	& \oplus \left(\begin{matrix}
	\zeta^{}_{3}\xi^{}_{2}-\zeta^{}_{2}\xi^{}_{3} \\ \zeta^{}_{2}\xi^{}_{1}-\zeta^{}_{1}\xi^{}_{2} \\
	\zeta^{}_{1}\xi^{}_{3}-\zeta^{}_{3}\xi^{}_{1}
	\end{matrix}\right)^{}_{\bf 3} \oplus \left(\begin{matrix}
	\zeta^{}_{3}\xi^{}_{3}-\zeta^{}_{2}\xi^{}_{2} \\
	\zeta^{}_{1}\xi^{}_{3}+\zeta^{}_{3}\xi^{}_{1} \\
	-\zeta^{}_{1}\xi^{}_{2}-\zeta^{}_{2}\xi^{}_{1}
	\end{matrix}\right)^{}_{\bf 3^{\prime}_{}} \; . \label{eq:33p}
	\end{align}
\end{itemize}
With the above decomposition rules and the assignments of relevant fields and modular forms, one can easily find out the Lagrangian invariant  under the modular $S^{}_4$ symmetry group.

As has been mentioned in Sec.~\ref{sec:modular}, there exist five linearly-independent modular forms of the lowest non-trivial weight $k^{}_{Y}=2$, denoted as $Y^{}_i(\tau)$ for $i = 1, 2, \cdots, 5$. They transform as a doublet ${\bf 2}$ and a triplet ${\bf 3}^{\prime}_{}$ under the $S^{}_4$ symmetry, namely~\cite{Penedo:2018nmg},
\begin{eqnarray}
Y^{}_{\bf 2}(\tau) \equiv \left(\begin{matrix} Y^{}_{1}(\tau) \\ Y^{}_2 (\tau)  \end{matrix}\right) \; , \quad Y^{}_{\bf 3^{\prime}_{}} (\tau) \equiv  \left(\begin{matrix} Y^{}_{3}(\tau) \\ Y^{}_4 (\tau) \\ Y^{}_{5} (\tau) \end{matrix}\right) \; .
\label{eq:S4Y1}
\end{eqnarray}
The expressions of modular forms can be derived with the help of the Dedekind $\eta$ function~\cite{Novichkov:2019sqv}
\begin{eqnarray}
\eta(\tau) \equiv q^{1/24}_{} \prod_{n=1}^{\infty}(1-q^{n}_{}) \; ,
\label{eq:eta}
\end{eqnarray}
with $q \equiv e^{2\pi {\rm i} \tau}_{}$, and its derivative~\cite{Penedo:2018nmg}
\begin{equation}
\begin{split}
Y(a^{}_{1}, \dots ,a^{}_{6}|\tau) \equiv
& \frac{\rm d}{\rm{d}\tau}\bigg[ a^{}_{1} \log \eta \left(\tau+\frac{1}{2} \right) + a^{}_{2} \log \eta \left(4\tau\right) + a^{}_{3} \log \eta \left(\frac{\tau}{4} \right) \\
& + a^{}_{4} \log \eta \left(\frac{\tau + 1}{4} \right) + a^{}_{5} \log \eta \left(\frac{\tau + 2}{4} \right) + a^{}_{6} \log \eta \left(\frac{\tau + 3 }{4} \right) \bigg] \; ,
\end{split}
\label{eq:geneY}
\end{equation}
with the coefficients $a^{}_i$ (for $i = 1, 2, \cdots, 6$) fulfilling $a^{}_{1}+ \cdots +a^{}_{6} = 0$. More explicitly, we have~\cite{Novichkov:2019sqv}
\begin{eqnarray}
Y^{}_{1}(\tau)  & \equiv & {\rm i} Y(1,1,-1/2,-1/2,-1/2,-1/2|\tau) \; , \nonumber \\
Y^{}_{2}(\tau)  & \equiv & {\rm i} Y(0,0,\sqrt{3}/2,-\sqrt{3}/2,\sqrt{3}/2,-\sqrt{3}/2|\tau) \; , \nonumber \\
Y^{}_{3}(\tau)  & \equiv & {\rm i} Y(1,-1,0,0,0,0|\tau) \; , \nonumber \\
Y^{}_{4}(\tau)  & \equiv & {\rm i} Y(0,0,-1/\sqrt{2},{\rm i}/\sqrt{2},1/\sqrt{2},-{\rm i}/\sqrt{2}|\tau) \; , \nonumber \\
Y^{}_{5}(\tau)  & \equiv & {\rm i} Y(0,0,-1/\sqrt{2},-\rm{i}/\sqrt{2},1/\sqrt{2},\rm{i}/\sqrt{2}|\tau) \; ,
\label{eq:Yexp}
\end{eqnarray}
which can be expanded as the Fourier series~\cite{Novichkov:2019sqv}, i.e.,
\begin{eqnarray}
Y^{}_{1}(\tau) &=& -3\pi\left(1/8 + 3q + 3q^{2}_{} + 12q^{3}_{} + 3q^{4}_{} + 18q^{5}_{} + 12q^{6}_{} + 24q^{7}_{} + 3q^{8}_{} + 39q^{9}_{} + \cdots\right) \; , \nonumber  \\
Y^{}_{2}(\tau) &=&  3\sqrt{3}\pi q^{1/2}_{}(1 + 4q + 6q^{2}_{} + 8q^{3}_{} + 13q^{4}_{} + 12q^{5}_{} + 14q^{6}_{} + 24q^{7}_{} + 18q^{8}_{} + \cdots) \; , \nonumber  \\
Y^{}_{3}(\tau) &=& \pi\left(1/4 - 2q + 6q^{2}_{} - 8q^{3}_{} + 6q^{4}_{} - 12q^{5}_{} + 24q^{6}_{} - 16q^{7}_{} + 6q^{8}_{} + 26q^{9}_{} + \cdots\right) \; , \label{eq:Y3q} \nonumber \\
Y^{}_{4}(\tau) &=& -\sqrt{2}\pi q^{1/4}_{}(1 + 6q + 13q^{2}_{} + 14q^{3}_{} + 18q^{4}_{} + 32q^{5}_{} + 31q^{6}_{} + 30q^{7}_{} + 48q^{8}_{} + \cdots) \; , \nonumber  \\
Y^{}_{5}(\tau) &=& -4\sqrt{2}\pi q^{3/4}_{}(1 + 2q + 3q^{2}_{} + 6q^{3}_{} + 5q^{4}_{} + 6q^{5}_{} + 10q^{6}_{} + 8q^{7}_{} + 12q^{8}_{}+\cdots) \; . 
\end{eqnarray}

\end{document}